\documentclass[final,3p,times,authoryear]{elsarticle}
\usepackage{graphicx}
\usepackage{booktabs}
\usepackage{url}
\usepackage{multicol}
\usepackage{acronym}
\usepackage{color}
\usepackage{xspace}
\usepackage{amsmath}
\usepackage{pifont}
\usepackage{longtable}
\usepackage{marginnote}
\newcommand{\cmark}{\ding{51}}%
\newcommand{\xmark}{\ding{55}}%

\newcommand{\figscale}{.55}

\newenvironment{fivenum}{\small%
  \begin{tabular}{llllll}
    \toprule
    & {\em Min.} & {\em 1st Qu.} & {\em Median} & {\em 3rd Qu.} & {\em Max.} \\   
    \midrule
}{%
  \bottomrule
  \end{tabular}
}

\acrodef{ANVUR}{National Agency for the Assessment of Universities and Research}
\acrodef{ASN}{National Scientific Qualification}
\acrodef{CI}{confidence interval}
\acrodef{MIUR}{Ministry of University and Research}
\acrodef{SD}{Scientific Discipline}

\acrodef{MCS}{Mathematics and Computer Sciences}
\acrodef{PHY}{Physics}
\acrodef{CHE}{Chemistry}
\acrodef{EAS}{Earth Sciences}
\acrodef{BIO}{Biology}
\acrodef{MED}{Medical Sciences}
\acrodef{AVM}{Agricultural Sciences and Veterinary Medicine}
\acrodef{CEA}{Civil Engineering and Architecture}
\acrodef{IIE}{Industrial and Information Engineering}
\acrodef{APL}{Antiquities, Philology, Literary Studies, Art History}
\acrodef{HPP}{History, Philosophy, Pedagogy and Psychology}
\acrodef{LAW}{Law}
\acrodef{ECS}{Economics and Statistics}
\acrodef{PSS}{Political and Social Sciences}
\acrodef{ERA}{Excellence in Research for Australia}

\begin{document}

\begin{frontmatter}

  \title{Assessing evaluation procedures for individual researchers:\\the case of the Italian National Scientific Qualification\tnoteref{titnote}}
\tnotetext[titnote]{The final version of this paper is published in the Journal of Informetrics. Please cite as: Moreno Marzolla, Assessing evaluation procedures for individual researchers: the case of the Italian National Scientific Qualification, Journal of Informetrics 10(2), May 2016, pp. 408-438, ISSN 1751-1577, DOI \url{http://dx.doi.org/10.1016/j.joi.2016.01.009}}
    
  \author{Moreno Marzolla\corref{cor1}}
  \ead{moreno.marzolla@unibo.it}
  \cortext[cor1]{Correspondence to: Moreno Marzolla, Department of Computer Science and Engineering, University of Bologna, mura Anteo Zamboni 7, I-40126 Bologna, Italy. Phone +39 051 20 94847, Fax +39 051 20 94510}
  \address{Department of Computer Science and Engineering, University of Bologna, Italy}

\begin{abstract}
  The Italian National Scientific Qualification~(ASN) was introduced
  as a prerequisite for applying for tenured associate or full
  professor positions at state-recognized universities. The~ASN is
  meant to attest that an individual has reached a suitable level of
  scientific maturity to apply for professorship positions. A five
  member panel, appointed for each scientific discipline, is in charge
  of evaluating applicants by means of quantitative indicators of
  impact and productivity, and through an assessment of their research
  profile. Many concerns were raised on the appropriateness of the
  evaluation criteria, and in particular on the use of bibliometrics
  for the evaluation of individual researchers. Additional concerns
  were related to the perceived poor quality of the final evaluation
  reports. In this paper we assess the~ASN in terms of appropriateness
  of the applied methodology, and the quality of the feedback provided
  to the applicants. We argue that the~ASN is not fully compliant with
  the best practices for the use of bibliometric indicators for the
  evaluation of individual researchers; moreover, the quality of final
  reports varies considerably across the panels, suggesting that
  measures should be put in place to prevent sloppy practices in
  future~ASN rounds.
\end{abstract}

\begin{keyword}
  National Scientific Qualification; ASN; Evaluation of individuals;
  Bibliometrics; Italy
\end{keyword}

\end{frontmatter}

\section{Introduction}\label{sec:introduction}

The~\acf{ASN} was introduced in 2010 as part of a global reform of the
Italian university system. The new rules require that applicants for
professorship positions in state-recognized universities must first
acquire a national scientific qualification for the discipline and
role applied to.

The~\ac{ASN} is to be held once a year; at the time of writing, two
rounds have been completed, started in 2012 and 2013,
respectively. Applicants are evaluated using quantitative indicators
as well as expert assessment. The Italian~\ac{MIUR} appoints 184
evaluation committees, one for each scientific discipline. Each
committee is made of five members: four are selected among full
professors from Italian universities, and one from foreign
universities or research institutions. Each committee processes all
applications for both the associate and full professor levels in its
field of competence.

Candidates are evaluated according to their scientific profile
(research output and other scientific titles, see
Section~\ref{sec:asn}). However, as an attempt to limit the unfair
selection practices that have been associated with the Italian
\emph{concorso}~\citep{gerosa01}, applicants are also evaluated
according to three bibliometric indicators of impact and scientific
productivity defined by the~\ac{MIUR}. The reliance of the~\ac{ASN} on
bibliometric indicators was welcome by part of the academic community
as a step towards more objective evaluation practices, but was also
heavily criticized by others as a form of ``career assessment by
numbers'' -- a term first used in~\cite{KellyJe06} -- and
against the best practices for the correct use of bibliometrics for
the evaluation of individual
researchers~\citep{banfi03}. Further complaints were
raised as soon as the final results were made available. The fraction
of qualified applicants varied considerably across~\acp{SD}, from a
minimum of 15.1\% to a maximum of 81.1\%~\citep{marzolla15}. Such
large differences can not be explained in terms of uncompetitive
applicants; rather, they suggest that the committees adopted different
criteria for qualification, if not unfair evaluation
practices~\citep{Abramo2015}.  In addition, many applicants perceived
the individual evaluations they received as hastily written and poorly
motivated.

The issues above are not specific to the~\ac{ASN}: indeed, defining
open, fair, and transparent evaluation procedures for career
advancement of scientists is a challenging task, as witnessed by the
plurality of hiring practices adopted in different
countries~\citep{Bennion2010,Brink2013,Dettmar2004,Vicker2006}.
The~\ac{ASN} is an interesting case study, since it produced a large
amount of data that have been made available on the Web for a short
period of time. The data include, for each applicant: the list of
publications and other scientific titles; the values of bibliometric
indicators; the outcome of the application (qualified/not qualified),
and a written assessment by the evaluation panel.

In this paper we address the following two questions: (\emph{i})~does
the~\ac{ASN} comply with the best practices for the use of
bibliometric indicators for evaluating individual researchers?
(\emph{ii})~do the final reports provide useful feedback to the
applicants? Both questions refer to the \emph{quality} of
the~\ac{ASN}, intended as its level of transparency and fairness.

The case study illustrated in this paper provides some important
lessons about the risks and unintended side-effects of evaluation
procedures for academics, especially when too much emphasis is put on
quantity rather than quality. As bibliometrics is used more and more
frequently to support hiring and promotion decisions~\citep{Sahel11},
it is important to share the experience gathered from the field so
that errors are not repeated. On top of that, national-wide research
evaluation campaigns such as the~\ac{ASN} face additional challenges
due to the large number of applications that must be processed. In
these situations it is tempting for evaluation committees to ``cut
corners'' and employ sloppy practices to speed up the evaluation
process, that reflect negatively on those being evaluated.

As valuable byproducts, we study the frequency of publication
categories appearing in the application forms, and the structure of
collaboration networks across scientific fields. The distribution of
publication types can be used to understand how researchers in
different disciplines disseminate their work. The investigation of the
structure and dynamics of inter-disciplinary research collaboration is
an important topic by itself that attracted considerable
interest~\citep{Newman2001,vanRijnsoever2011,Wagner2011, Abbasi2012},
and is important, e.g., for funding agencies to identify and possibly
support joint research and development activities.

\paragraph{Related work} 

Hiring and promotion procedures for academic staff vary considerably
across countries. The Academic Career Observatory from the European
University Institute published a comprehensive overview of the
recruiting and career advancement procedures in European countries and
abroad\footnote{\url{http://www.eui.eu/ProgrammesAndFellowships/AcademicCareersObservatory/Index.aspx},
  accessed on 2015-10-06.}, including information on salaries, access
to non-nationals and gender issues.

Qualification procedures somewhat similar to the~\ac{ASN} are already
in place in other European countries, like Germany, France, and
Spain. In Germany there are two paths towards professorship positions:
Assistants working towards the \emph{Habilitation}, and Junior
Professors that must carry out a variety of tasks (including research,
teaching, management) but are not required to get the
Habilitation. The German Habilitation is essentially a second PhD, and
may consist of either a thesis, or several publications of high
quality~\citep{Enders2001}. Similarly, the French \emph{habilitation
  \`a diriger des recherches} is awarded to applicants with a strong
publication record over a period of years, and is required to
supervise PhD students and to apply to professor
positions~\citep{Musselin2004}. Finally, Spain introduced the
\emph{accreditation}~\footnote{\url{http://www.aneca.es/eng/Programmes/PEP},
  accessed on 2015-10-03} as a prerequisite to apply to
\emph{Agregat} and \emph{Catedr\`atic} positions (roughly equivalent
to associate and full professor). The accreditation is granted by the
Spanish national evaluation agency~(ANECA) after detailed assessment
of the applicant CV, including teaching, research experience, and list
of publications. Of the three procedures above, the Spanish
accreditation is the most similar to the~\ac{ASN}. However,
the~\ac{ASN} is, to the best of our knowledge, the only scientific
qualification that explicitly relies on bibliometric indicators of
scientific productivity and impact to evaluate applicants. Also, while
teaching activities play a significant role in the Spanish
accreditation, they are barely considered by the~\ac{ASN}
(see~\ref{app:statistics}).

A quantitative account of the~\ac{ASN} is given by~\cite{marzolla15}:
the author computes a set of descriptive statistics, showing among
other things the fraction of qualified applicants, and the
distribution of the values of bibliometric indicators. The study shows
that the fraction of successful applicants varies considerably
across~\acp{SD}, suggesting that the qualification criteria were
interpreted differently by each evaluation panel. This is confirmed by
the comparison of bibliometric indicators of qualified and not
qualified applicants, showing that some panels were more likely to
deviate from purely quantitative considerations for granting or
denying qualification. \cite{Abramo2015} examine the relationship of
the~\ac{ASN} outcome with the scientific merit of applicants, in order
to identify possible cases of discrimination or
favoritism. Discrimination refers to skilled (according to their
bibliometric indicators) applicants that are denied qualification,
while favoritism refers to under-performing applicants that are
granted qualification. The results reveal that applicants that are not
already employed by an academic institution (``outsiders'') tend to be
more penalized. Finally, \cite{pautasso2015} studies the proportions
and success rates of female applicants across the various~\acp{SD} to
investigate gender issues. While in most disciplines the success rates
of female applicants are comparable to that of male candidates, the
study observes a significantly lower proportion of female scientists
applying to most~\acp{SD}, especially for the full professor role.

\paragraph{Organization of this paper}
This paper is organized as follows. In Section~\ref{sec:asn} we give
some information on the~\ac{ASN}. In Section~\ref{sec:results} we
examine the evaluation forms: we study their length and average
similarity as proxies of their perceived quality. In
Section~\ref{sec:discussion} we discuss whether the~\ac{ASN}
methodology follows the current best practices for the correct use of
bibliometric indicators for the evaluation of researchers. Finally,
conclusions are presented in Section~\ref{sec:conclusions}.  Some
interesting descriptive statistics on the~\ac{ASN} dataset that have
been produced as a byproduct of the main analysis are described
in~\ref{app:statistics}.

\section{Background}\label{sec:asn}

In this section we provide some background on the~\ac{ASN} and the
Italian university system; for an historical perspective,
see~\cite{DegliEsposti2010}. 

In Italy, each professor and researcher is bound to a~\ac{SD}
representing a specific field of study. There are~184~\acp{SD}
organized in 14 areas shown in Table~\ref{tab:n-forms}. Each~\ac{SD}
is identified by a four-character code of the form $AA/MC$ where $AA$
is the numeric ID of the area (01--14), $M$ is a single letter
identifying the macro-sector, and $C$ is a single digit identifying
the discipline within the macro-sector. The full list can be found
in~\ref{app:list-sd}.

Before 2010, there were three tenured roles at Italian universities:
assistant professor (\emph{ricercatore universitario}), associate
professor (\emph{professore associato}) and full professor
(\emph{professore ordinario}). Hiring procedures were handled by
universities advertising the position, according to centrally-defined
rules mandated by state laws. Applicants had to undergo a written
and/or oral examination (\emph{concorso}) whose exact details differed
for each role.

Law 240/2010 replaced the role of tenured assistant professor with two
fixed-term positions, called \emph{Type A} and \emph{Type B}
researcher. Type B positions are supposed to be a path towards the
associate professor role, since universities hiring Type B researchers
must allocate funding for promotion in advance.  Under the new rules,
to apply for a permanent professor positions at any state-recognized
university, one has to first obtain the~\ac{ASN} in the same~\ac{SD}
and role (associate or full professor) applied for. A five-member
evaluation panel, appointed by the~\ac{MIUR} for each discipline,
grants or denies qualification after assessing the scientific profiles
of applicants.  The evaluation must take into account both the
qualitative and quantitative scientific profile of candidates. The
\emph{qualitative} profile consists of the list of publications and
other scientific titles, such as coordination of research projects,
patents, visiting positions at foreign institutions, and so on (the
teaching activity is not considered, though); each panel must also
provide an opinion on a limited set of publications submitted by each
applicant in full text. The \emph{quantitative} profile is assessed
using three numeric indicators of impact and productivity.

Two sets of indicators are defined: \emph{bibliometric} and
\emph{non-bibliometric} indicators.  Bibliometric indicators apply to
disciplines such as the hard sciences, biology and medicine, for which
``sufficiently complete'' citation databases exist.  Specifically,
bibliometric indicators apply to all disciplines of the nine
areas~\ac{MCS}, \ac{PHY}, \ac{CHE}, \ac{EAS}, \ac{BIO}, \ac{MED},
\ac{AVM}, \ac{CEA} and \ac{IIE}, except 08/C1--\emph{Design and
  technological planning of architecture}, 08/D1--\emph{Architectural
  design}, 08/E1--\emph{Drawing}, 08/E2--\emph{Architectural
  restoration and history} and 08/F1--\emph{Urban and landscape
  planning and design}, but including the whole macro sector
11/E--\emph{Psychology}.
  
The bibliometric indicators are the following (the normalization
procedure will be described shortly):
  
\begin{itemize}
\item[B.1] normalized number of journal papers;
\item[B.2] normalized number of citations received;
\item[B.3] normalized $h$-index.
\end{itemize}

Non-bibliometric indicators apply to all other disciplines (in
general, social sciences and humanities), and are:

\begin{itemize}
\item[N.1] normalized number of authored books;
\item[N.2] normalized number of journal papers and book chapters;
\item[N.3] normalized number of papers published on ``top'' journals.
\end{itemize}

The lists of ``top'' journals mentioned in N.3 have been defined by
panels of experts from the relevant~\acp{SD}, appointed by
the~\acf{ANVUR}, a public entity under control of~\ac{MIUR}.

Normalization of the raw indicators\footnote{ANVUR (2013), National
  Scientific Qualification -- normalization of indicators by
  academic age (\emph{Abilitazione scientifica nazionale -- la
    normalizzazione degli indicatori per l'et{\`a} accademica}),
  \url{http://www.anvur.org/attachments/article/253/normalizzazione_indicatori_0.pdf},
  accessed on 2015-10-06.} is used to limit the bias against
young applicants, and is based on the concept of \emph{scientific
  age}: the scientific age $\mathit{SA}(A)$ of applicant $A$ that
published the first paper in year $t_0(A)$ is defined as:

\begin{equation*}
  \mathit{SA}(A) := \max\left\{10, (2012 - t_0(A) + 1)\right\}
\end{equation*}

Indicators B.1, N.1, N.2 and N.3 are normalized by multiplying their
raw value by $10 / \mathit{SA}(A)$. Indicator B.2 is normalized by
dividing the raw number of citations by the scientific age. Finally,
the value of B.3 is computed from the normalized number of citations
per paper. Specifically, given a paper $p$, published in year $t_p$,
that at time $t \geq t_p$ has received $C(p, t)$ citations, the
normalized number of citations $S(p, t)$ for $p$ is defined as:

\begin{equation*}
  S(p, t) := \frac{4}{t - t_p + 1} C(p, t)
\end{equation*}

The normalized h-index $h_c$ is then the maximum integer such that
$h_c$ papers of a given applicant received at least $h_c$
\emph{normalized} citations each~\citep{hc-index}.

We remark that the terms \emph{bibliometric} and
\emph{non-bibliometric} are used in the official~\ac{MIUR}
documentation, although their meaning does not match the one used by
the scientometric community. For this reason we will use the generic
term ``quantitative indicator'' to refer to both bibliometric and
non-bibliometric indicators.

The values of quantitative indicators are compared to minimum
thresholds, defined as the medians of the values of the same
indicators for tenured professors of the same role and~\ac{SD} applied
for. Both the medians and of the values of quantitative indicators for
each applicant are computed by~\ac{ANVUR} using data from Scopus and
Web of Science (WoS). The list of publications used to compute the
medians, and the quantitative indicators of tenured professors, have
not been made publicly available, so the computations can not be
independently verified.

Under the initial interpretation of the~\ac{ASN} rules, qualification
could be granted only to applicants that strictly exceed at least two
(one, for non-bibliometric disciplines) medians; this was understood
to be a necessary but not sufficient condition for
qualification. Later, \ac{MIUR} relaxed this interpretation by
allowing panels to grant qualification also to applicants that do not
satisfy the constraint above, provided that such decision is
motivated\footnote{F. Profumo, Newsletter of the ministry of
  education, university and research concerning some aspects of the
  new discipline for granting the national scientific qualification
  introduced by law 240 on Dec. 30, 2010 (Newsletter of the Ministry
  of Education, University and Research concerning some aspects of the
  new discipline for acquiring the national scientific qualification
  introduced with Law 30 December 2010, n. 240 (\emph{Nota Circolare
    del Ministero dell'Istruzione, dell'Universit\`a e della Ricerca
    su alcuni aspetti della nuova disciplina per il conseguimento
    dell'abilitazione scientifica nazionale introdotta dalla legge 30
    dicembre 2010, n. 240}), January 11, 2013,
  \url{http://www.anvur.org/attachments/article/252/Circolare%20Profumo%20ASN%20gennaio%202013.pdf},
    accessed on 2015-10-06.}}. Applicants who failed to get the
qualification were prevented from applying again during the next two
years.\medskip

\begin{table}[tT]
    \centering%
  \begin{tabular*}{\textwidth}{@{\extracolsep{\fill}}lllrrr}
    \toprule
    {\em Id} & {\em Code} & {\em Area Name} & {\em Applications} & {\em Sample size} & {\em Coverage} \\
    \midrule
   1 & MCS & Mathematics and Computer Sciences & 2492 & 2116 & 84.91\% \\ 
    2 & PHY & Physics & 4372 & 4372 & 100.00\% \\ 
    3 & CHE & Chemistry & 2344 & 2344 & 100.00\% \\ 
    4 & EAS & Earth Sciences & 1231 & 1231 & 100.00\% \\ 
    5 & BIO & Biology & 6244 & 6244 & 100.00\% \\ 
    6 & MED & Medical Sciences & 9987 & 9266 & 92.78\% \\ 
    7 & AVM & Agricultural Sciences and Veterinary Medicine & 2093 & 1895 & 90.54\% \\ 
    8 & CEA & Civil Engineering and Architecture & 3599 & 3284 & 91.25\% \\ 
    9 & IIE & Industrial and Information Engineering & 4535 & 3860 & 85.12\% \\ 
   10 & APL & Antiquities, Philology, Literary Studies, Art History & 6324 & 6322 & 99.97\% \\ 
   11 & HPP & History, Philosophy, Pedagogy and Psychology & 5909 & 3975 & 67.27\% \\ 
   12 & LAW & Law & 3037 & 2774 & 91.34\% \\ 
   13 & ECS & Economics and Statistics & 4853 & 4848 & 99.90\% \\ 
   14 & PSS & Political and Social Sciences & 2129 & 1274 & 59.84\% \\     \midrule
    &&& $59149$ & $53805$ & $90.97\%$ \\
    \bottomrule
  \end{tabular*}
\caption{The table reports, for each area, the total number of
  submitted qualification applications (\emph{Applications}) and the
  number (\emph{Sample size}) and percentages (\emph{Coverage}) of
  applications for which the CV and evaluation forms have been
  collected.}\label{tab:n-forms}
\end{table}

\begin{figure}[tT]
  \centering%
  \includegraphics[width=.8\textwidth]{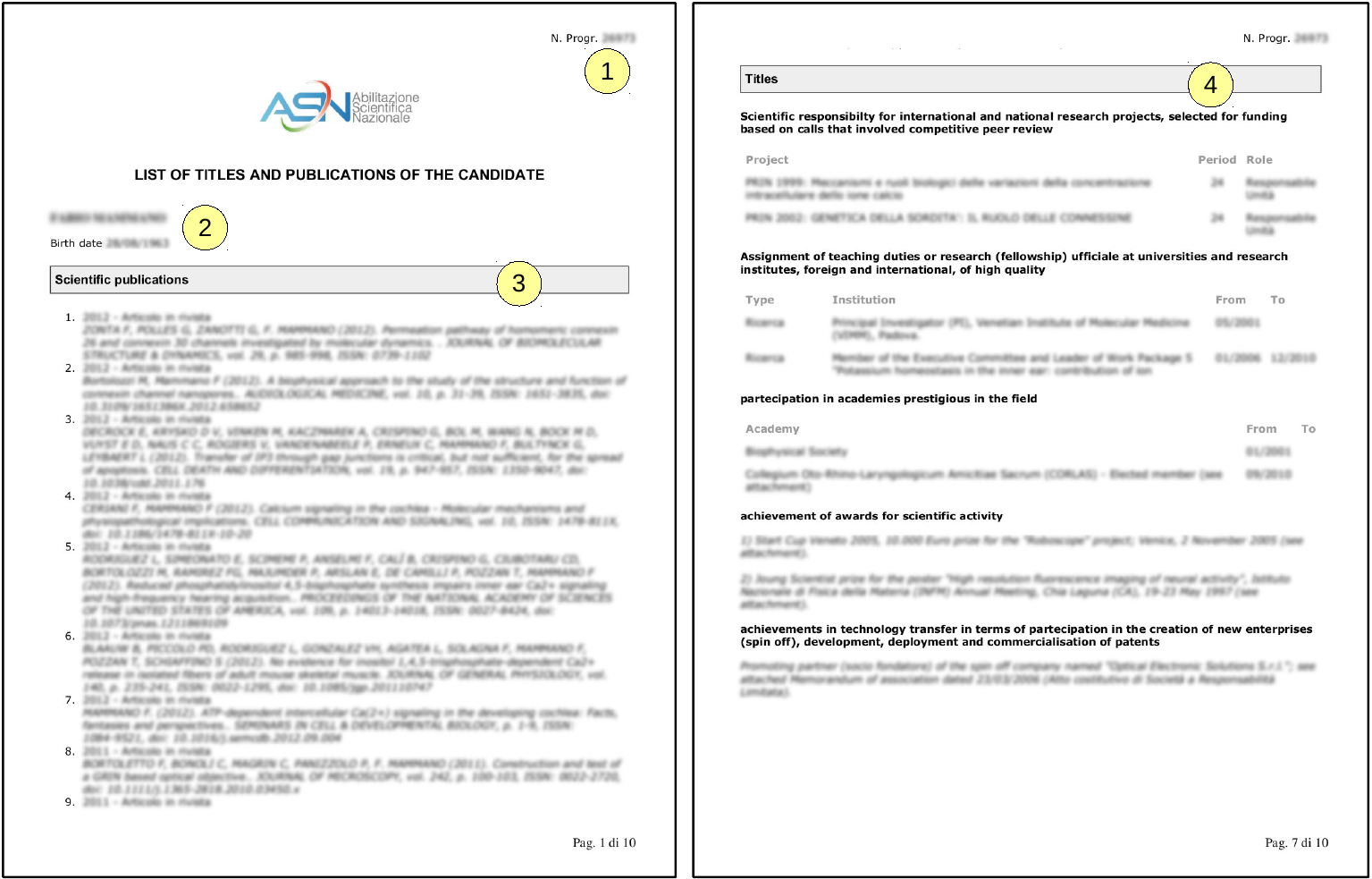}
  \caption{Structure of an application
    form.}\label{fig:application-form}
\end{figure}

All~\ac{ASN} applications were submitted electronically through the
Web site \url{http://abilitazione.miur.it}; each application was then
automatically converted to a PDF document, like the one shown in
Figure~\ref{fig:application-form}. The form contains the following
elements:

\begin{enumerate}
\item Unique application ID;
\item Applicant first, last name, and date of birth; the date of birth
  is a valuable detail because the triplet \emph{first name},
  \emph{last name}, and \emph{date of birth} is a robust unique
  identifier~\citep{smalheiser11};
\item List of publications;
\item List of additional scientific qualifications and titles.
\end{enumerate}

The values of quantitative indicators, the application forms, and the
final evaluations have been made publicly available for a short period
of time at the~\ac{ASN} Web site.  Table~\ref{tab:n-forms} shows the
number of submitted applications for each area, and the number of
application forms and final reports that have been collected and will
be analyzed in this paper. Our dataset includes $53,805$ pairs of
forms (for each applicant, we either managed to get both the
application and final report, or none of them). This corresponds to
about 90\% of all application forms, representing a sufficiently large
subset. Unfortunately, the coverage is not uniform across the
scientific areas: from Table~\ref{tab:n-forms} we observe that the
dataset is complete for areas~\ac{PHY}, \ac{CHE}, \ac{EAS}
and~\ac{BIO}. Areas~\ac{HPP} and~\ac{PSS} are only partially covered,
and no reports at all are available for the following
$14$ \acp{SD}:

\begin{itemize}
\item  01/A4--\emph{Mathematical physics} 
\item  06/D3--\emph{Blood diseases, oncology and rheumatology} 
\item  06/E1--\emph{Heart, thoracic and vascular surgery} 
\item  07/H1--\emph{Veterinary anatomy and physiology} 
\item  07/H5--\emph{Clinical veterinary surgery and obstetrics} 
\item  08/A1--\emph{Hydraulics, hydrology, hydraulic and marine constructions} 
\item  09/H1--\emph{Information processing systems} 
\item  11/A1--\emph{Medieval history} 
\item  11/A3--\emph{Contemporary history} 
\item  11/A4--\emph{Science of books and documents, history of religions} 
\item  11/C2--\emph{Logic, history and philosophy of science} 
\item  11/C4--\emph{Aesthetics and philosophy of languages} 
\item  12/B1--\emph{Business, navigation and air law} 
\item  14/C1--\emph{General and political sociology, sociology of law} \end{itemize}

We remark that the coverage refers to the fraction of applications for
which the PDF forms have been collected; the values of the
quantitative indicators for all applicants have been collected, and
where the subject of the analysis in~\citep{marzolla15}.

It is interesting to observe that each application form has a unique
ID that appears to have been generated sequentially. There are gaps in
the sequence of IDs; these gaps can be attributed to the fact that our
sample is not complete, to applications that have been created but not
finalized, and to applications that have been withdrawn after
submission. The maximum ID in our dataset is
$94765$, much larger than the number of applications
($59,149$, see~\cite{marzolla15}). The \emph{German tank
  problem}~\citep{ruggles47} technique can be used to get an accurate
estimate of the total number of applications. A 95\%~\ac{CI} is
$[94765.04, 94771.5]$, which is compatible with the
rough estimate using the maximum ID alone.

\ref{app:statistics} provides additional descriptive statistics of
the~\ac{ASN} dataset.

\section{Analysis of final reports}\label{sec:results}

\begin{figure}[tT]
  \centering%
  \includegraphics[width=.8\textwidth]{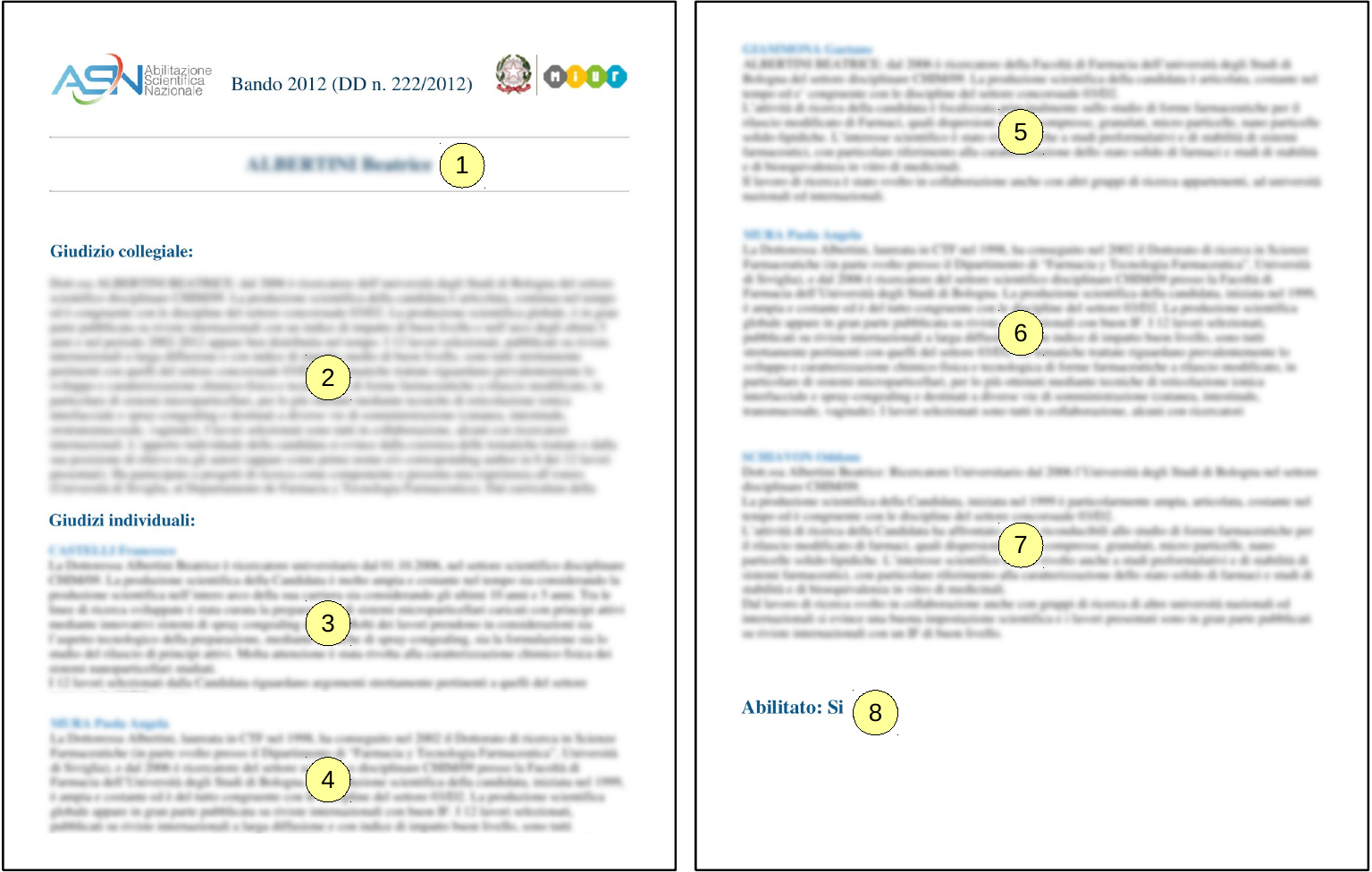}
  \caption{Structure of a final report.}\label{fig:evaluation-form}
\end{figure}

In this section we focus our attention on the final reports containing
the assessment of each applicant. A typical report is shown in
Figure~\ref{fig:evaluation-form}, and contains the following elements:

\begin{enumerate}
\item[1.] Applicant's last and first name;
\item[2.] Collegial assessment (\emph{Giudizio collegiale}) formulated
  by the whole panel;
\item[3--7.] Individual assessment (\emph{Giudizi individuali})
  formulated by each member of the evaluation committee; the name of
  the committee member is indicated above the evaluation, that are
  therefore not anonymous;
\item[8.] Result (qualified / not qualified).
\end{enumerate}

Most of the final reports are written in Italian, with the possible
exception of the evaluations written by the foreign panel members.
However, a few panels used a different language for the whole report.

The reports are extremely important, especially for applicants who
failed to get qualification: in these cases, it is reasonable to
expect that the reports motivate the decision for denying
qualification, and provide feedback to improve the quality of the
applicant research output. A good report should list the strengths
and weaknesses of each applicant, and provide an evaluation on each
paper submitted in full text: does the paper address a topic that
falls within the aim and scope of the~\ac{SD}?  is the contribution
significant? is the publication type appropriate? did the publication
produce an impact on the scientific community? This is not dissimilar
to the feedback that authors of a scientific paper submitted to
peer-review expect to receive~\citep{Shashok2008}.
    
Unfortunately, as soon as the reports started to be made available,
complaints were raised about their perceived poor quality. Among
others, two issues were frequently reported: (\emph{i})~very short
reports that do not provide any useful feedback; (\emph{ii})~reports
that are very similar across applicants for the same~\ac{SD}, as if
they were based on a template with only minor modifications. These
issues are examples of \emph{anti-patterns}~\citep{koenig95}, i.e.,
common but counterproductive solutions to some problem.

The task of deciding whether a report is appropriate can not be fully
automated, since this would require natural language processing
capabilities far beyond the current state of the art; besides, the
definition of ``appropriate'' is subjective and can not be encoded in
any formal rule. However, the two anti-patterns above can be
identified with the help of simple text metrics. In the following we
focus on the length of the reports and their dissimilarity, measured
through a suitable \emph{text distance} function.

\subsection{Length of final reports}

\begin{figure}[tT]
  \centering\includegraphics[width=.9\textwidth]{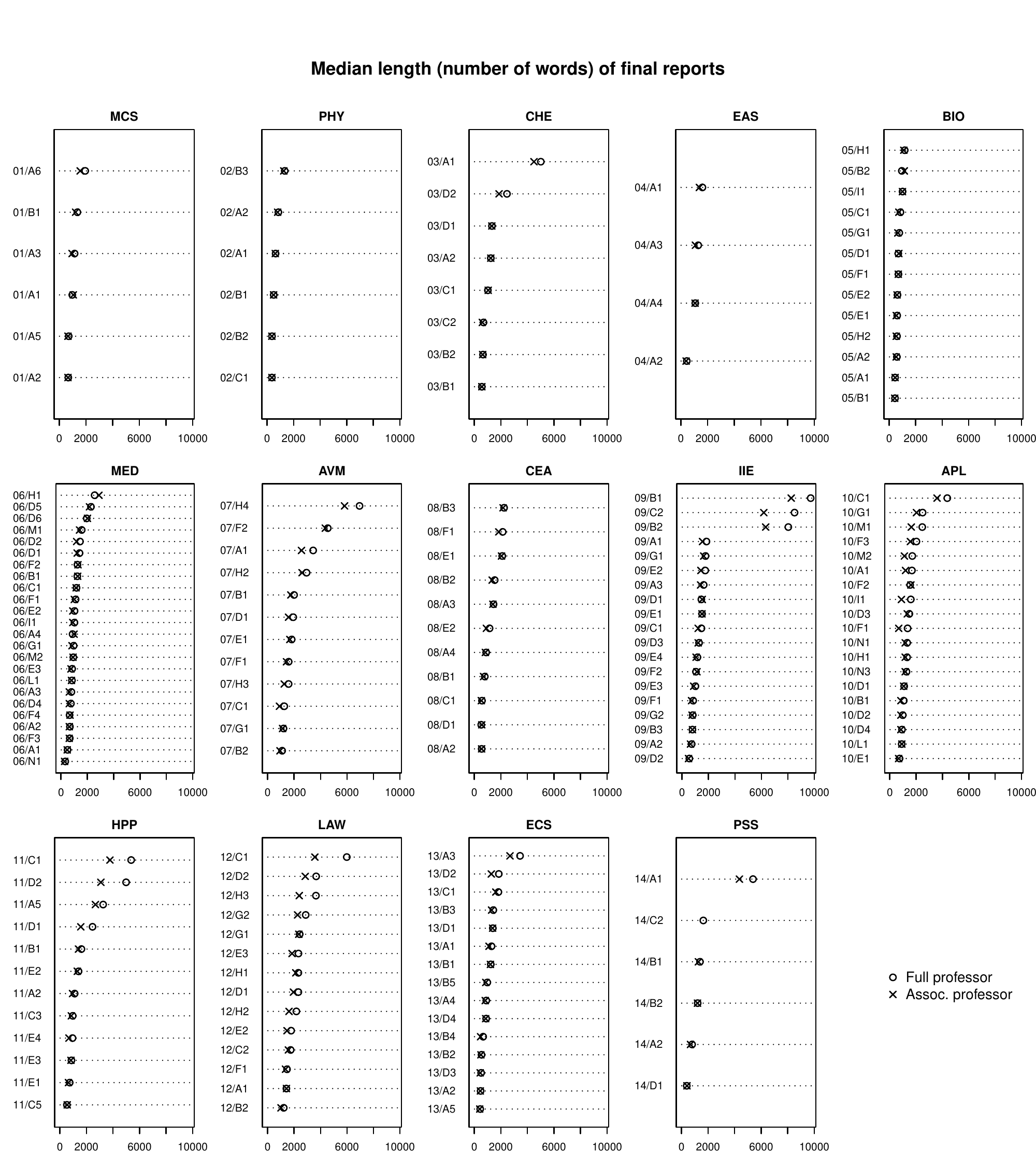}
  \caption{Median length (number of words) of final reports for each
    discipline and role.}\label{fig:length}
\end{figure}

\begin{table}[tT]
  \centering%
  \begin{fivenum}
    Associate & 153 & 616 & 936 & 1342 & 10030\\
    Full & 185 & 658 & 1050 & 1481 & 10970\\
  \end{fivenum}
  \caption{Five number summary for the length (number of words) of
    final reports.}\label{tab:length}
\end{table}

The length of final reports is the number of characters or words they
contain; we use the number of words as a matter of convenience, since
this allows us to deal with smaller numbers that are more easy to
grasp intuitively. 

Figure~\ref{fig:length} shows the median length of the final reports
for each discipline in our dataset; Table~\ref{tab:length} shows the
five number summary~\citep{tukey77} of all lengths. The medians for
full and associate professor applications are both about 1000 words,
corresponding roughly to two pages like those shown in
Figure~\ref{fig:evaluation-form}. However, there are also a
significant number of very short reports (200-300 words or less). They
may be appropriate in some circumstances, e.g., if the applicant is
obviously under-qualified, or has applied to an unrelated~\ac{SD}: in
these cases there is no need to provide a lengthy
explanation. Figure~\ref{fig:length}, however, shows that there are
panels that systematically produced shorter reports, and this can not
be explained by occasional low-quality candidates.

\begin{figure}[tT]
\fbox{%
  \begin{minipage}[t]{\textwidth}
      \textbf{\color{blue} Collegial evaluation}\smallskip
      
      The scientific production of the applicant lies in the area of
      \emph{AAA BBB CCC}, shows good coherence with the scientific
      discipline and good continuity, but is of limited quality. The
      applicant took part to national and international research
      projects.\bigskip
      
      \textbf{\color{blue} Individual evaluations}\medskip
            
      \emph{\color{blue} PANEL MEMBER XXX}\smallskip
     
      Publications: the applicant presents publications related to
      \emph{AAA BBB CCC}; good fit with this discipline and temporal
      continuity; quality is poor and international visibility is very
      poor. Scientific titles: the applicant took part to national and
      international research projects. The applicant is not
      qualified.\medskip
      
      \emph{(Four other similar individual evaluations omitted)}
    \end{minipage}}
\caption{A fragment of an actual report (translation
  from the original in Italian)}\label{fig:short-eval}
\end{figure}

As an actual example, Figure~\ref{fig:short-eval} shows the English
translation of a portion of one of the short reports (about 300 words)
for an applicant who failed to get qualification; we only show the
collegial evaluation and one of the individual assessments, the other
four being very similar.  As can be seen, the content is quite vague:
the publications are considered of ``limited quality'', and the
international visibility ``very poor'', without any further
explanation. Such evaluation is far from useful, and does not provide
any of the feedback mentioned at the beginning of this section.

As a general rule, short reports should be closely scrutinized since
they are likely to be of low quality, such as the one above. However,
long reports should not be blindly considered better. For example,
some panels listed the publications provided in full text by the
applicant; in some cases the list appears multiple times in the same
report, i.e., in the collegial assessment and in the five individual
evaluations. The mere fact of listing the same publications over and
over again increases the length but does not improve the quality of
the evaluation, unless the lists are used to provide an assessment of
each publication, as is actually done by some panels (e.g., the
reports of 09/B1--\emph{Manufacturing technology and systems} provide
a detailed evaluation on each publication submitted for
consideration).  We will show later on how the length of final reports
should be combined with their textual distance to obtain a less
fragile quality indicator.

\begin{table}[tT]
  \centering%
  \begin{fivenum}
    $RD$ Full prof. applications & 0 & 0.003 & 0.009 & 0.021 & 0.239\\
    $RD$ Assoc. prof. applications & 0 & 0.005 & 0.012 & 0.023 & 0.237\\
  \end{fivenum}
  \caption{Five number summary for the relative difference of the
    average length of final reports for qualified and not qualified
    applicants.}\label{tab:rd}
\end{table}

To study whether there are significant differences between the average
lengths of reports for qualified and not qualified applicants, we
define the following quantities. Let $\mathit{LQ}_i$ be the average
length of reports for qualified applicants in discipline $i$, and
$\mathit{LNQ}_i$ the average length of reports for \emph{not}
qualified applicants in $i$.  The relative difference $\mathit{RD}_i$
of the lengths is defined as:
  
\begin{equation*}
  \mathit{RD}_i := \frac{| \mathit{LQ}_i - \mathit{LNQ}_i |}{\max(\mathit{LQ}_i, \mathit{LNQ}_i)}
\end{equation*} 
  
Table~\ref{tab:rd} shows the five number summary of $\mathit{RD}_i$
for full and associate professor applications, respectively. The 3rd
quartile is about 0.02 for both roles; this means that the relative
difference between reports for successful and unsuccessful
applications is very small, less than 2\% in 75\% of the
disciplines.

\begin{figure}[tT]
  \centering\includegraphics[scale=\figscale]{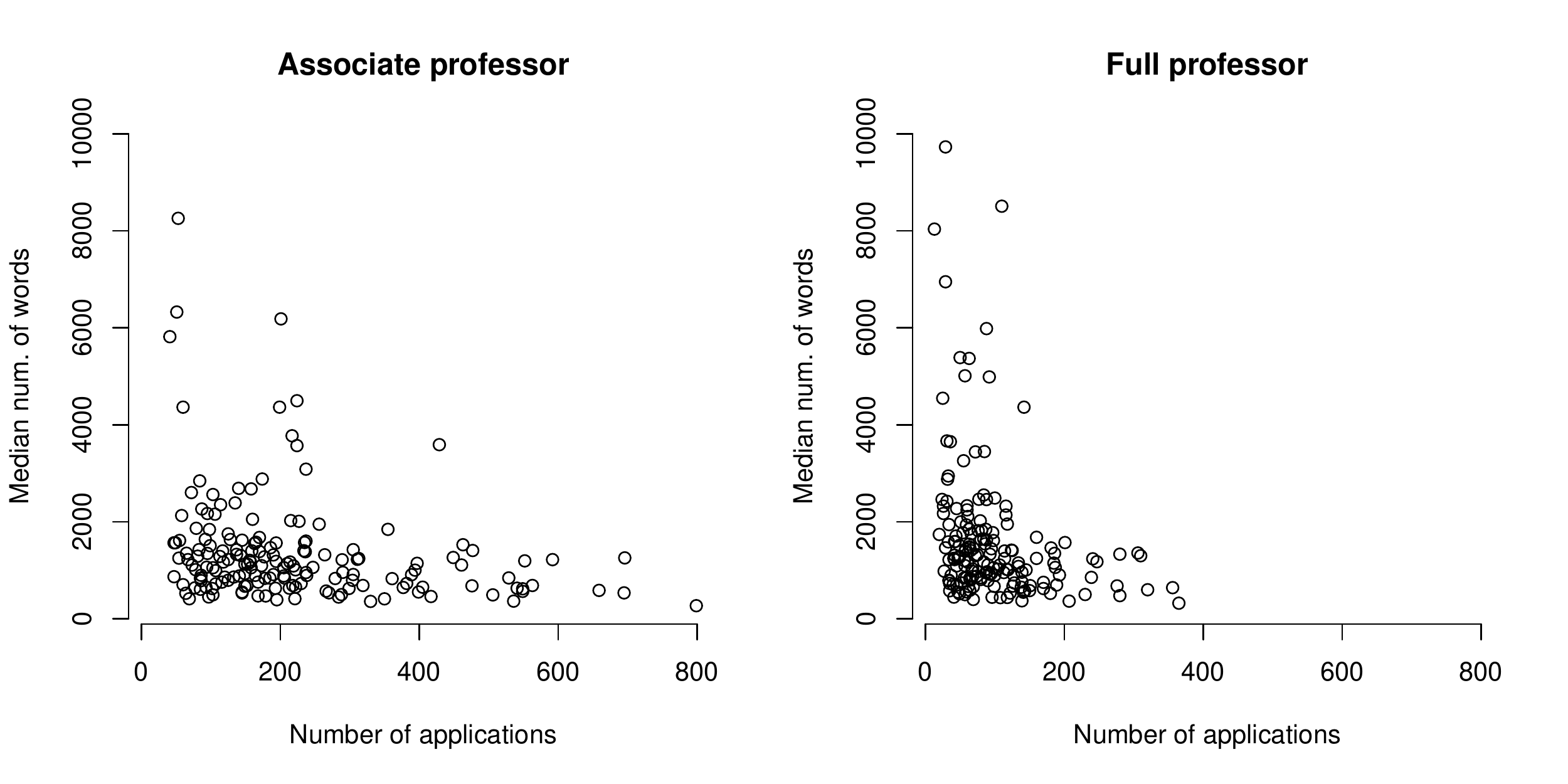}
  \caption{Correlation between the number of applications and the
    median length of the final reports}\label{fig:cor-length}
\end{figure}

We observe negative correlation between the median length of
evaluations and the number of applications in each~\ac{SD}
(Figure~\ref{fig:cor-length}). The rank order correlation coefficient
is~$\rho=-0.29$ with 95\%~\ac{CI}
$[-0.43, -0.14]$ for associate professor,
and~$\rho=-0.35$ with 95\%~\ac{CI}
$[-0.48, -0.20]$ for full professor applications. The
negative correlation may be explained by the fact that the panels that
had to process more applications could dedicate less time to each
one. However, the correlation is weak, so we can not rule out the
possibility that the lengths are unrelated to the number of
applications.

\subsection{Similarity among evaluation forms}\label{sec:text-distance}

\begin{figure}[tT]
  \begin{small}
    \setlength{\columnseprule}{1pt}
    \begin{multicols}{2}
      \sf
      \textbf{Applicant $A$}\medskip
      
      The publications presented by the applicant are considered
      \textbf{sufficiently consistent} with the scope of discipline
      XX/XX or the related interdisciplinary topics. The evaluation of
      the scientific contribution of the publications, in relation to
      the scope of scientific discipline XX/XX, is assessed using
      parameter set 1 in Annex B of the minutes of the meeting held on
      X XXXX 2013 describing the criteria adopted by the panel, is
      \textbf{good}. The productivity of the applicant, assessed on
      the basis of publications submitted in relation to discipline
      XX/XX, with particular reference to the last five years prior to
      the call, using the parameter set 2 in Annex B of the minutes of
      the meeting held on XX XXXX 2013 describing the criteria adopted
      by the panel, is overall good. Other qualifications submitted by
      the applicant to support his authority and scientific maturity
      in relation to scientific discipline XX/XX, are considered,
      based on the parameter set 3 described in Annex B of the minutes
      of the meeting held on XX XXXX 2013 which describes the criteria
      adopted by the panel, \textbf{excellent}.

      \columnbreak

      \textbf{Applicant $B$}\medskip
      
      The publications presented by the applicant are considered
      \textbf{consistent} with the scope of discipline XX/XX or the
      related interdisciplinary topics.  The evaluation of the
      scientific contribution of the publications, in relation to the
      scope of scientific discipline XX/XX, is assessed using
      parameter set 1 in Annex B of the minutes of the meeting held on
      X XXXX 2013 describing the criteria adopted by the panel, is
      \textbf{fair}. The productivity of the applicant, assessed on
      the basis of publications submitted in relation to discipline
      XX/XX, with particular reference to the last five years prior to
      the call, using the parameter set 2 in Annex B of the minutes of
      the meeting held on XX XXXX 2013 describing the criteria adopted
      by the panel, is overall good. Other qualifications submitted by
      the applicant to support his authority and scientific maturity
      in relation to scientific discipline XX/XX, are considered,
      based on the parameter set 3 described in Annex B of the minutes
      of the meeting held on XX XXXX 2013 which describes the criteria
      adopted by the panel, \textbf{good}.
    \end{multicols}    
  \end{small}
  \caption{Two assessments written by the same member of one
    evaluation panel on two applicants (translation by the
    author). The differences are reported in bold}\label{fig:reports}
\end{figure}

Another problem that has been observed in some~\acp{SD} is that the
evaluations are almost identical, as if they were variations of the
same template. To illustrate the problem, we report in
Figure~\ref{fig:reports} the translation of two actual evaluations
written by the same committee member for two applicants, $A$ (who got
the qualification) and $B$ (who was denied qualification). The
differences between the two texts consists of the three words shown in
bold. From these tiny differences it is difficult to understand why
applicant $A$ was granted qualification but $B$ was not: indeed, the
terms ``consistent'', ``fair'' and ``good'' bears a positive meaning,
suggesting that $B$ met all the criteria for qualification.  The
practice of ``cloning'' the evaluations to change just a few words is
a sloppy practice that reduces the quality of final reports. In the
following we assess the extent of this practice in all~\acp{SD}.

We measure the similarity among the reports of each~\ac{SD} by
computing the \emph{text distance} among documents. Two families of
text distances are used in the literature: \emph{semantic distances},
that measure whether two documents contains the same information, and
\emph{string distances}, that measure the similarity of their
syntactic representation. String distances have the advantage of being
easy to compute and content-agnostic; furthermore, they provide a
stronger evidence that two documents share a common textual template,
as in the example above.

The \emph{Levenshtein distance}~\citep{levenshtein65} measures the
similarity of two documents as the minimum number of edit operations
required to transform one document into the other (see
\ref{sec:levenshtein} for details). We use the \emph{normalized}
Levenshtein distance that produces a value in the interval $[0, 1]$.
A distance of~0 denotes that the two documents are identical, while~1
denotes that the documents have no character in common. In practice,
the normalized Levenshtein distance rarely exceeds~0.8 even between
unrelated documents written in different languages; higher values are
therefore very unlikely to be observed.

Given $N$ reports $\{R_1, \ldots, R_N\}$ for a given~\ac{SD} and role,
we compute the pairwise distances $L_{ij}$ between document $R_i$ and
$R_j$ for all $1 \leq i < j \leq N$. We strip all non-alphanumeric
characters and translate uppercase letters to lowercase, to make the
distance robust against changes in formatting marks. The empirical
distribution of $L_{ij}$ provides information about the mutual
similarity of the documents in the set. Since the computation of all
distances is time consuming, we consider a random sample of $N = 100$
reports for each~\ac{SD} and role.

\begin{figure}[tT]
  \centering\includegraphics[width=.9\textwidth]{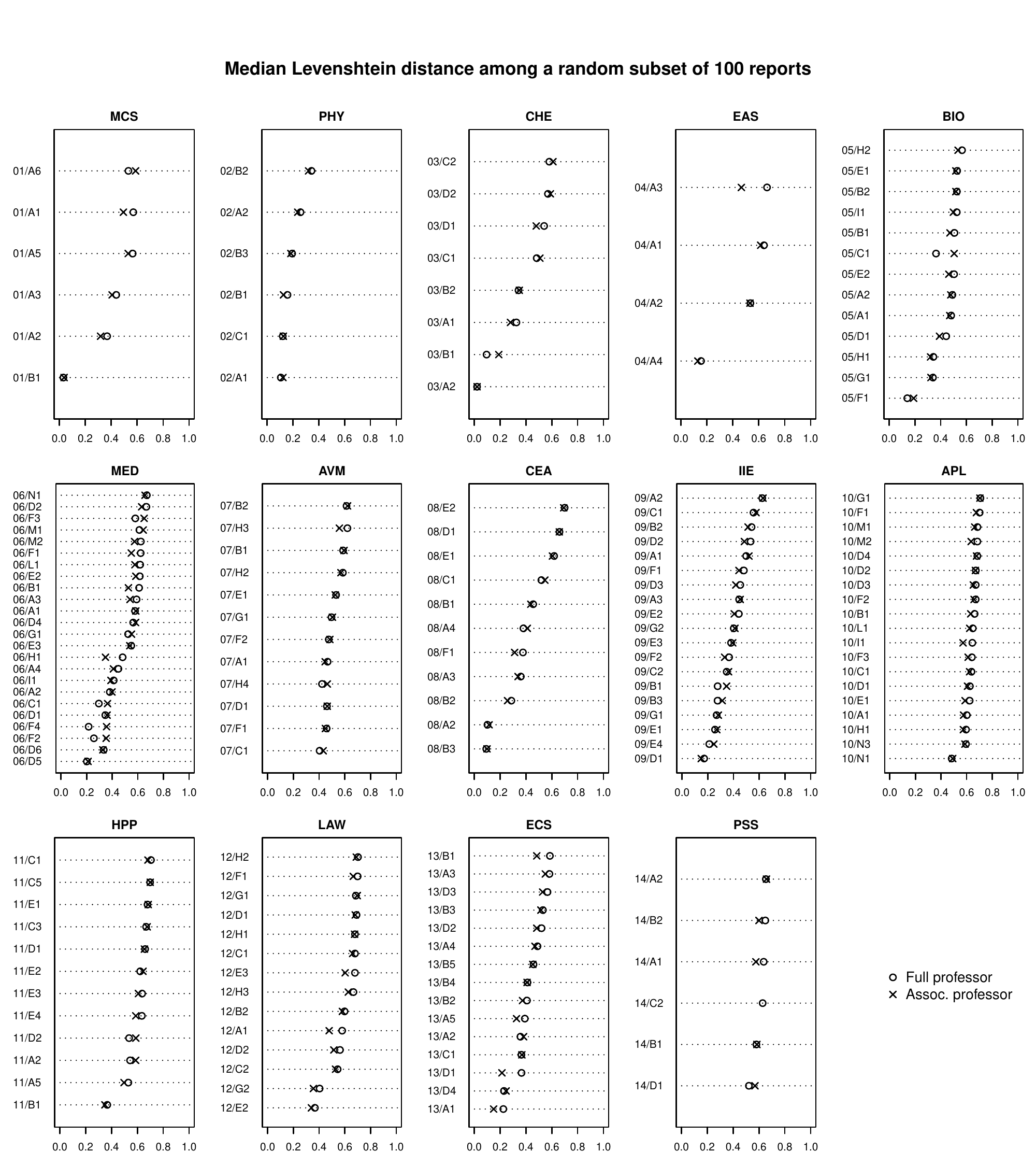}
  \caption{Median of the normalized Levenshtein distance among a
    random sample of 100 reports, for each~\ac{SD} and role; higher is
    better.}\label{fig:report-dist}
\end{figure}

Figure~\ref{fig:report-dist} shows the medians of the normalized
Levenshtein distances among the reports in the samples, for each
discipline and role. Low values are a clear indication of low quality
reports that are similar each other. On the other hand, high values
can not be automatically considered an indication of better reports.
As an example, let us consider~\ac{SD} 06/N1--\emph{Applied medical
  technologies}. According to Figure~\ref{fig:report-dist}, its final
reports have the higher distance within area~\ac{MED};
Figure~\ref{fig:length}, however, shows that the reports are, on
average, the shortest in~\ac{MED}. Manual examination of the reports
shows that they are indeed short and uninformative. The problem here
is that two short documents that differ in a few words have higher
distance than two long documents that differ in the exact same words
(see~\ref{sec:levenshtein} for a technical explanation). Therefore,
short documents are more likely to have higher normalized distance
than longer ones.

\begin{figure}[tT]
  \centering\includegraphics[width=.9\textwidth]{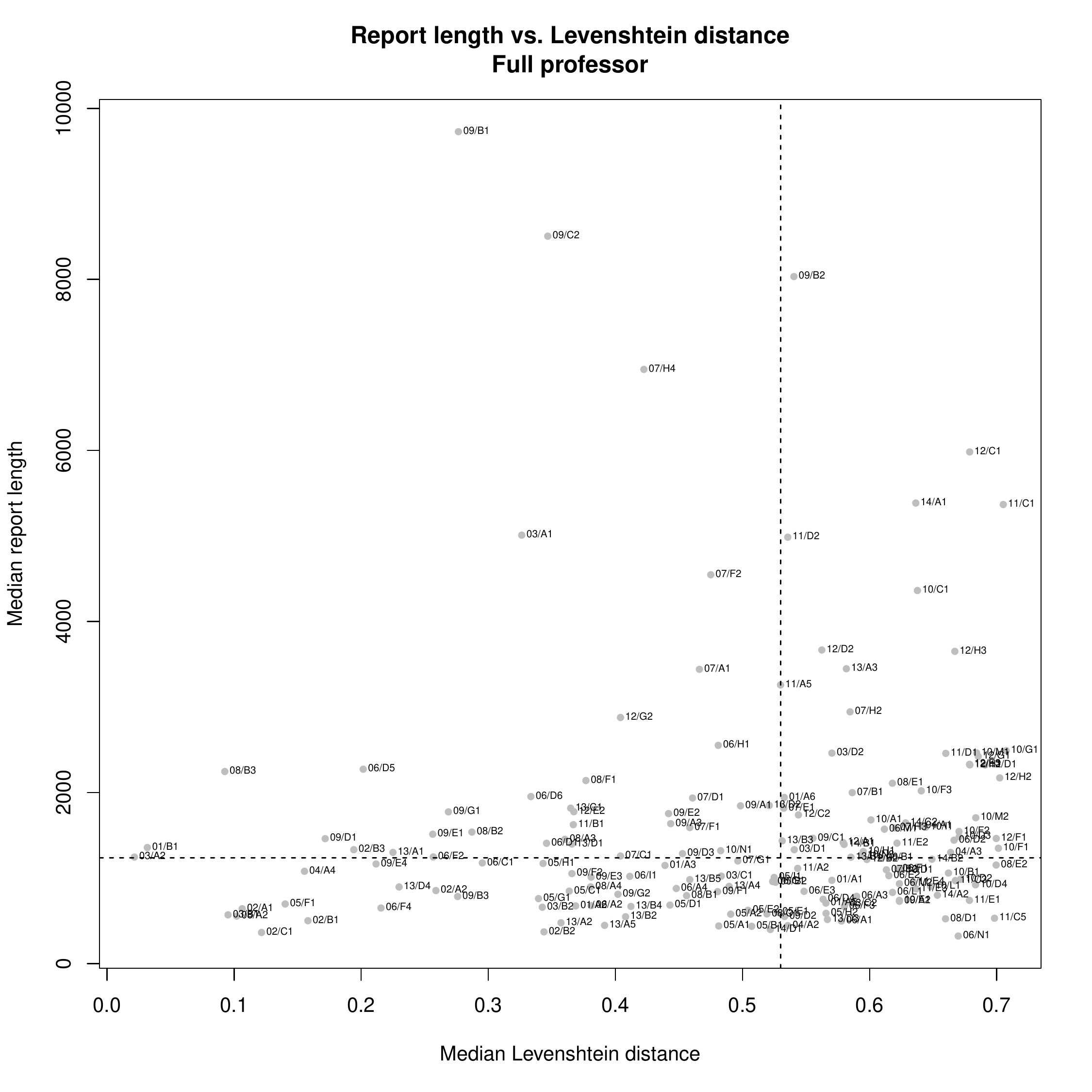}
  \caption{Median report length versus normalized Levenshtein distance
    between final reports for the full professor applications. The
    dashed lines denote the median length and
    distance.}\label{fig:scatter-dist-len}
\end{figure}

The discussion above suggests that the length and normalized textual
distance, if taken alone, are only weak indicators of the quality of
the final reports since they can produce false positives: low values
are clear indication of poorly written reports, but higher values do
not automatically denote better ones. A more robust indicator can be
obtained by jointly considering both metrics. A simple way to do so is
to produce a scatter plot such as the one in
Figure~\ref{fig:scatter-dist-len}, where data points
represent~\acp{SD} whose coordinates are the median distance and the
median report length, respectively; the dashed lines in the figure
correspond to the global median length and distance. The plot for the
associate professor level is almost identical and is not shown.
  
The ``good'' reports are those that are both long and with high
pairwise normalized distance, that are located in the upper right
portion of the scatter plot. ``Bad'' reports, that are both short and
undifferentiated, are located in the lower left portion. Hence, the
scatter plot provides an easy way to identify the~\acp{SD} that more
likely produced low quality reports.

\section{Discussion}\label{sec:discussion}

\begin{table}[tT]
  \centering
  \begin{tabular}{rlll}
    \toprule
    \multicolumn{2}{l}{\em Criterion} & {\em Pass} & {\em Fail} \\
    \midrule
    1.  & Quantitative evaluation should support qualitative, expert assessment & \cmark & \\
    2.  & Measure performance against the research missions of the institution, group or researcher & & \xmark\\
    3.  & Protect excellence in locally relevant research & \cmark & \\
    4.  & Keep data collection and analytic processes open, transparent and simple & & \xmark \\
    5.  & Allow those evaluated to verify data and analysis & & \xmark\\
    6.  & Account for variation by field in publication and citation practices & \cmark & \\
    7.  & Base assessment of individual researchers on a qualitative judgment of their portfolio & \cmark & \\
    8.  & Avoid misplaced concreteness and false precision & & \xmark\\
    9.  & Recognize the systemic effects of assessment and indicators & & \xmark\\
    10. & Scrutinize indicators regularly and update them & \cmark &\\
    \bottomrule
  \end{tabular}
  \caption{Ten criteria proposed in the Leiden manifesto for research
    metrics. See text for discussion.}\label{tab:best-practices}
\end{table}

In the previous section we have analyzed whether the~\ac{ASN} results
provide useful feedback to the applicants. In this section we take a
broader view by discussing the appropriateness of the~\ac{ASN}
methodology, including the use of bibliometric indicators to evaluate
individual applicants. Indeed, the~\ac{ASN} is the only national
scientific qualification procedure that also uses quantitative
indicators of productivity and impact for assessing applicants.

The recently published \emph{Leiden manifesto for research
  metrics}~\citep{Hicks2015} describes ten best practices that should
be followed when using bibliometrics as a tool to evaluate individuals
or organizations. The best practices are quite general and can be
applied to any scientific discipline; it is therefore instructive to
understand whether the~\ac{ASN} complies with them.  Since the best
practices are provided as high-level requirements rather than formal
rules, the discussion will be somewhat subjective; to substantiate our
claims we will refer to the quantitative analysis from the previous
section, whenever appropriate. The best practices from the Leiden
manifesto are the following:

\begin{enumerate}

\item \textbf{Quantitative evaluation should support qualitative,
  expert assessment.} In the~\ac{ASN}, a five member panel is
  appointed for each~\ac{SD}, and must take into account both the
  quantitative and qualitative profile of applicants. Indeed,
  \cite{marzolla15} observed that there is a considerable fraction of
  applicants that satisfies the quantitative requirements but is
  denied qualification; this fraction is not homogeneous across
  the~\acp{SD}, suggesting that the qualitative assessment was carried
  out differently. Anyway, this denotes that the~\ac{ASN} is -- at
  least in principle -- not driven by the numbers only, and therefore
  this requirement appears to be met.

\item \textbf{Performance should be measured against the research
  missions of the institution, group or researcher.} The~\ac{ASN}
  rules have been centrally defined and applied to all~\acp{SD}, with
  the only distinction between bibliometric and non-bibliometric
  disciplines (see Section~\ref{sec:asn}). The quantitative indicators
  put in place for the two classes of disciplines are certainly not
  enough to cope with the variability of research practices and goals
  across fields of study. While each panel had the possibility to
  override at least part of the rules, very few of them did so. The
  Leiden manifesto remarks that ``no single evaluation model applies
  to all contexts''; unfortunately this is precisely what happened
  with the~\ac{ASN}.
    
\item \textbf{Excellence in locally relevant research should be
  protected.} Research excellence should not be identified with
  English-language publications only, since that would penalize the
  activities that have regional or national scope (typical of social
  sciences and humanities). The~\ac{ASN} relies on bibliometric data
  from Scopus and Web of Science for bibliometric disciplines, where
  English is used the most anyway. Social sciences and humanities use
  paper-counting metrics and lists of ``top'' journals for each
  specific field. These journals are published in a variety of
  languages, allowing locally relevant research to be recognized.

\item \textbf{The data collection and analytical processes should be
  kept open, transparent and simple.} The~\ac{ASN} is based on a new
  and unproven methodology that has not been discussed with the
  scientific community, nor has been validated by experts in research
  evaluation. The official documents do not contain any reference to
  the state of the art and to the known best practices. Therefore we
  can conclude that the~\ac{ASN} does not provide a suitable level of
  openness and transparency.

\item \textbf{Allow those evaluated to verify data and analysis.}
  The~\ac{ASN} fails (badly) to meet this requirement.  In principle,
  applicants could verify the values of their quantitative indicators
  by computing them using data from Scopus and WoS. However, not
  everyone has access to these databases; furthermore, the values can
  be updated by the providers without notice, and therefore there is
  no guarantee that the values observed by the applicants at some time
  are the same values that are made available to the panels.  The
  situation concerning the medians is worse: the list of publications
  used to compute them has not been made public, and it is therefore
  impossible to verify that the medians are correct. It should be
  observed that~\ac{ANVUR} released an updated set of threshold
  values\footnote{Consiglio direttivo ANVUR, On the computation of
    medians for the national scientific qualification (\emph{Sul
      calcolo delle mediane per l'abilitazione nazionale}), Sep. 14,
    2012,
    \url{http://www.anvur.it/attachments/article/253/mediane_spiegate_definitivo_14_settembre_2012.pdf}}
  to fix errors that were discovered after publication of the initial
  set of thresholds. This raises the serious concern that other issues
  may have gone unnoticed.
      
\item \textbf{Account for variation by field in publication and
  citation practices.} It is well known that citation-based metrics
  vary significantly across fields of
  study~\cite{AlbarranCrOrRu11}. Publication practices also vary:
  Table~\ref{tab:top-pubtypes} in the Appendix lists the four most
  frequent publication types for each~\ac{SD} in our dataset, showing
  considerable differences also among disciplines within the same
  macro-sector. The~\ac{ASN} addressed these issues by defining
  different thresholds for each~\ac{SD} and role. Provisions were also
  made to cope with multimodal distributions of quantitative
  indicators caused by the coexistence of different scientific
  communities within the same field of study.
    
\item \textbf{Base assessment of individual researchers on a
  qualitative judgment of their portfolio.} The~\ac{ASN} complies with
  this requirement. Indeed, applicants were required to submit a
  selection of their best publications to the evaluation panel. The
  quality of those publications had to be assessed as part of the
  applicant evaluation. Note, however, that the analysis of the final
  reports described in Section~\ref{sec:results} questions the
  accuracy of the qualitative judgment of applicants on some~\acp{SD}.
    
\item \textbf{Avoid misplaced concreteness and false precision.} The
  thresholds of the quantitative indicators used in the~\ac{ASN} were
  supposed to be ``hard'' values that had to be strictly exceeded by
  applicants to be considered for qualification. This neglects the
  fact that the indicators are subject to uncertainties: should an
  applicant with contemporary $h$-index equal to $10.4$ be rejected if
  the minimum threshold is $10.5$? While a few panels recognized the
  problem and adopted less stringent requirements, the vast majority
  stuck with the simplistic interpretation of the hard thresholds.
    
\item \textbf{Recognize the systemic effects of assessment and
  indicators.} Scientists that are evaluated according to a set of
  rules inevitably tend to optimize their behavior to better fit the
  rules. The Leiden manifesto suggests that a pool of different
  metrics should be preferred to a single metric that can be easily
  gamed. The~\ac{ASN} complies with this suggestion, since it bases
  the evaluation on three quantitative indicators. However, we have
  observed that the values of the indicators are positively correlated
  in bibliometric disciplines~\citep{marzolla15}, suggesting that in
  fact they might measure the same thing.  This suggests that the
  systemic effects of indicators were not properly dealt with.
    
\item \textbf{Scrutinize indicators regularly and update them.}
  The~\ac{MIUR} made explicit provision to revise the criteria and
  parameters every five years\footnote{Ministerial Decree 76/2012,
    Criteria and Parameters for evaluation of applicants for the
    National Scientific Qualification (\emph{Regolamento recante
      criteri e parametri per la valutazione dei candidati ai fini
      dell'attribuzione dell'Abilitazione Scientifica Nazionale}),
    Ministerial Decree 76, June 7, 2012,
    \url{http://www.anvur.org/attachments/article/251/dm_07_06_12_regolamento_abilitazione.pdf},
    art. 9}.
\end{enumerate}

The discussion above is summarized in Table~\ref{tab:best-practices},
where we show whether each requirement from the Leiden manifesto is
satisfied or not. Since the~\ac{ASN} was defined before the
publication of the manifesto, it is unreasonable to expect that
the~\ac{ASN} fully complies. However, the manifesto did not appear out
of the blue: the issues associated with the use of bibliometrics to
evaluate individuals are well known and have already been described in
the literature~\citep{proper,LaloeMo09,Sahel11,ieee-bib,oecd-bib}.

\section{Conclusions}\label{sec:conclusions}

In this paper we have considered the Italian~\ac{ASN} as a case study
in the evaluation of individual researchers for promotion. In
particular, we were interested in assessing the appropriateness of
the~\ac{ASN} in terms of fairness and quality of feedback provided to
applicants. To do so, we addressed the following two questions:
(\emph{i})~does the~\ac{ASN} comply with the best practices for the
use of bibliometric indicators for evaluating individual researchers?
(\emph{ii})~do the final reports provide useful feedback to the
applicants?

The answer is partially positive for question~(\emph{i}). We have
considered the ten best practices for evaluating individual
researchers through bibliometrics, according to the Leiden manifesto
for research metrics. The~\ac{ASN} fails to satisfy five out of then
requirements: the metrics are defined without taking into
consideration the mission of the institution, group or researcher; the
data collection and analysis process is not transparent; applicants
are unable to verify the data and analysis; the possible lack of
precision of the quantitative indicators used is not taken into
consideration; finally, the systemic effect of the assessment is
overlooked.

To answer question (\emph{ii}) we have used two simple measures
(length and normalized Levenshtein distance) to analyze the content of
the individual reports containing a written assessment of each
applicant. These measures, both in isolation and in combination, show
that the perceived poor quality of some reports is indeed justified.

Our analysis of the Italian~\ac{ASN} highlights several issues, listed
below in no particular order:

\begin{enumerate}

\item \textbf{Understand and follow best practices.} Rules and
  procedures for evaluating individual researchers should be defined
  with the help of experts in research evaluation, and should be
  discussed and accepted by the scientific communities. In the case of
  the~\ac{ASN}, \cite{marzolla15} observed that the definition of the
  quantitative indicators and their medians generated several
  unintended side effects, including the ``paradox of academic
  twins''\footnote{\url{http://www.roars.it/online/sulla-revisione-dellasn-alcune-proposte/},
    accessed on 2015-10-06.}  (an applicant with a proper subset of
  the publications of another one might have \emph{higher} -- i.e.,
  better -- quantitative indicators). Also, in some disciplines the
  thresholds for qualification at the associate level were higher than
  those for the full professor level, implying that in those
  disciplines there are higher requirements for the lower academic
  rank. Finally, the use of journal rankings presents known
  issues~\citep{Vanclay2011} that have not been addressed in the list
  of top journals used in non-bibliometric disciplines.

\item \textbf{Allocate enough resources.} Nation-wide research
  evaluation procedures should expect to receive a large number of
  applications; it is therefore important that sufficient resources
  (time and manpower) are allocated so that all applications are
  evaluated fairly and accurately. Some evaluation panels of
  the~\ac{ASN} were subject to unrealistic deadlines, and therefore
  required multiple extensions that delayed publications of the
  results. This issue could be addressed by splitting the workload of
  the same~\ac{SD} across multiple panels and/or simplifying the
  qualification procedure in such a way that the workload becomes
  manageable.
  
\item \textbf{Check for common anti-patterns.} An obvious corollary of
  the point above is that when evaluation panels are subject to
  unrealistic deadlines they inevitably tend to work sloppily in order
  to save time. A frequent complaint on the~\ac{ASN} refers to the
  poor quality of the final reports. The analysis in
  Section~\ref{sec:results} shows that those complaints are in some
  cases justified. Suitable quality assurance mechanisms are put in
  place to improve the quality of final reports and provide consistent
  feedback to applicants; such mechanisms are already being used in
  some conferences to improve the quality of the paper review
  process~\citep{Canfora15}.
  
\item \textbf{Be transparent.} Transparency is an important deterrent
  against unfair practices and corruption. In this context,
  transparency means that the output of the evaluation process should
  be public, so that \emph{ex-post} analyses can be performed to
  identify issues. Moreover, if bibliometrics is used as part of the
  evaluation process, the indicators and their values should be
  verifiable by applicants.
  
\end{enumerate}

\section*{Acknowledgments}

The author thanks Giuseppe De Nicolao for providing feedback on a
preliminary version of this analysis, and the online community of
\emph{Redazione ROARS} (\url{http://www.roars.it/}) for valuable ideas
and discussion.

\appendix

\section{List of Scientific Disciplines}\label{app:list-sd}

The list below enumerates all scientific areas (first indentation
level), macro-sectors (second indentation level) and scientific
disciplines.

\begin{multicols}{2}
\begin{scriptsize}
\begin{description}\item[01] Mathematics and computer sciences
\begin{description}
\item[01/A] Mathematics
\begin{description}
\item[01/A1] Mathematical logic, mathematics education and history of mathematics
\item[01/A2] Geometry and algebra
\item[01/A3] Mathematical analysis, probability and statistics
\item[01/A4] Mathematical physics
\item[01/A5] Numerical analysis
\item[01/A6] Operational research
\end{description}
\item[01/B] Computer Science
\begin{description}
\item[01/B1] Computer Science
\end{description}
\end{description}
\item[02] Physics
\begin{description}
\item[02/A] Physics of fundamental interactions
\begin{description}
\item[02/A1] Experimental physics of fundamental interactions
\item[02/A2] Theoretical physics of fundamental interactions
\end{description}
\item[02/B] Physics of matter
\begin{description}
\item[02/B1] Experimental physics of matter
\item[02/B2] Theoretical physics of matter
\item[02/B3] Applied physics
\end{description}
\item[02/C] Astronomy, astrophysics, Earth and planetary physics
\begin{description}
\item[02/C1] Astronomy, astrophysics, Earth and planetary physics
\end{description}
\end{description}
\item[03] Chemistry
\begin{description}
\item[03/A] Analytical and physical chemistry
\begin{description}
\item[03/A1] Analytical chemistry
\item[03/A2] Models and methods for chemistry
\end{description}
\item[03/B] Inorganic chemistry and applied technologies
\begin{description}
\item[03/B1] Principles of chemistry and inorganic systems
\item[03/B2] Chemical basis of technology applications
\end{description}
\item[03/C] Organic, industrial and applied chemistry
\begin{description}
\item[03/C1] Organic chemistry
\item[03/C2] Industrial and applied chemistry
\end{description}
\item[03/D] Medicinal and food chemistry and applied technologies
\begin{description}
\item[03/D1] Medicinal, toxicological and nutritional chemistry and applied technologies
\item[03/D2] Drug technology, socioeconomics and regulations
\end{description}
\end{description}
\item[04] Earth sciences
\begin{description}
\item[04/A] Earth sciences
\begin{description}
\item[04/A1] Geochemistry, mineralogy, petrology, volcanology, Earth resources and applications
\item[04/A2] Structural geology, stratigraphy, sedimentology and paleontology
\item[04/A3] Applied geology, physical geography and geomorphology
\item[04/A4] Geophysics
\end{description}
\end{description}
\item[05] Biology
\begin{description}
\item[05/A] Plant biology
\begin{description}
\item[05/A1] Botany
\item[05/A2] Plant physiology
\end{description}
\item[05/B] Animal biology and anthropology
\begin{description}
\item[05/B1] Zoology and anthropology
\item[05/B2] Comparative anatomy and cytology
\end{description}
\item[05/C] Ecology
\begin{description}
\item[05/C1] Ecology
\end{description}
\item[05/D] Physiology
\begin{description}
\item[05/D1] Physiology
\end{description}
\item[05/E] Experimental and clinical biochemistry and molecular biology
\begin{description}
\item[05/E1] General biochemistry and clinical biochemistry
\item[05/E2] Molecular biology
\end{description}
\item[05/F] Experimental biology
\begin{description}
\item[05/F1] Experimental biology
\end{description}
\item[05/G] Experimental and clinical pharmacology
\begin{description}
\item[05/G1] Pharmacology, clinical pharmacology and pharmacognosy
\end{description}
\item[05/H] Human anatomy and histology
\begin{description}
\item[05/H1] Human anatomy
\item[05/H2] Histology
\end{description}
\item[05/I] Genetics and microbiology
\begin{description}
\item[05/I1] Genetics and microbiology
\end{description}
\end{description}
\item[06] Medicine
\begin{description}
\item[06/A] Pathology and laboratory medicine
\begin{description}
\item[06/A1] Medical genetics
\item[06/A2] Experimental medicine, pathophysiology and clinical pathology
\item[06/A3] Microbiology and clinical microbiology
\item[06/A4] Pathology
\end{description}
\item[06/B] General clinical medicine
\begin{description}
\item[06/B1] Internal medicine
\end{description}
\item[06/C] General clinical surgery
\begin{description}
\item[06/C1] General surgery
\end{description}
\item[06/D] Specialized clinical medicine
\begin{description}
\item[06/D1] Cardiovascular and respiratory diseases
\item[06/D2] Endocrinology, nephrology, food and wellness sciences
\item[06/D3] Blood diseases, oncology and rheumatology
\item[06/D4] Skin, contagious and gastrointestinal diseases
\item[06/D5] Psychiatry
\item[06/D6] Neurology
\end{description}
\item[06/E] Specialized clinical surgery
\begin{description}
\item[06/E1] Heart, thoracic and vascular surgery
\item[06/E2] Plastic and paediatric surgery and urology
\item[06/E3] Neurosurgery and maxillofacial surgery
\end{description}
\item[06/F] Integrated clinical surgery
\begin{description}
\item[06/F1] Odontostomatologic diseases
\item[06/F2] Eye diseases
\item[06/F3] Otorhinolaryngology and audiology
\item[06/F4] Musculoskeletal diseases and physical and rehabilitation medicine
\end{description}
\item[06/G] Paediatrics
\begin{description}
\item[06/G1] Paediatrics and child neuropsychiatry
\end{description}
\item[06/H] Gynaecology
\begin{description}
\item[06/H1] Obstetrics and gynecology
\end{description}
\item[06/I] Radiology
\begin{description}
\item[06/I1] Diagnostic imaging, radiotherapy and neuroradiology
\end{description}
\item[06/L] Anaesthesiology
\begin{description}
\item[06/L1] Anaesthesiology
\end{description}
\item[06/M] Public health
\begin{description}
\item[06/M1] Hygiene, public health, nursing and medical statistics
\item[06/M2] Forensic and occupational medicine
\end{description}
\item[06/N] Applied medical technologies
\begin{description}
\item[06/N1] Applied medical technologies
\end{description}
\end{description}
\item[07] Agricultural and veterinary sciences
\begin{description}
\item[07/A] Agricultural economics and appraisal
\begin{description}
\item[07/A1] Agricultural economics and appraisal
\end{description}
\item[07/B] Agricultural and forest systems
\begin{description}
\item[07/B1] Agronomy and field, vegetable, ornamental cropping systems
\item[07/B2] Arboriculture and forest systems
\end{description}
\item[07/C] Agricultural, forest and biosytems engineering
\begin{description}
\item[07/C1] Agricultural, forest and biosystems engineering
\end{description}
\item[07/D] Plant pathology and entomology
\begin{description}
\item[07/D1] Plant pathology and entomology
\end{description}
\item[07/E] Agricultural chemistry and agricultural genetics
\begin{description}
\item[07/E1] Agricultural chemistry, agricultural genetics and pedology
\end{description}
\item[07/F] Food technology and agricultural microbiology
\begin{description}
\item[07/F1] Food science and technology
\item[07/F2] Agricultural microbiology
\end{description}
\item[07/G] Animal science and technology
\begin{description}
\item[07/G1] Animal science and technology
\end{description}
\item[07/H] Veterinary medicine
\begin{description}
\item[07/H1] Veterinary anatomy and physiology
\item[07/H2] Veterinary pathology and inspection of foods of animal origin
\item[07/H3] Infectious and parasitic animal diseases
\item[07/H4] Clinical veterinary medicine and pharmacology
\item[07/H5] Clinical veterinary surgery and obstetrics
\end{description}
\end{description}
\item[08] Civil engineering and architecture
\begin{description}
\item[08/A] Landscape and infrastructural engineering
\begin{description}
\item[08/A1] Hydraulics, hydrology, hydraulic and marine constructions
\item[08/A2] Sanitary and environmental engineering, hydrocarbons and underground fluids, safety and protection engineering
\item[08/A3] Infrastructural and transportation engineering, real estate appraisal and investment valuation
\item[08/A4] Geomatics
\end{description}
\item[08/B] Structural and geotechnical engineering
\begin{description}
\item[08/B1] Geotechnics
\item[08/B2] Structural mechanics
\item[08/B3] Structural engineering
\end{description}
\item[08/C] Design and technological planning of architecture
\begin{description}
\item[08/C1] Design and technological planning of architecture
\end{description}
\item[08/D] Architectural design
\begin{description}
\item[08/D1] Architectural design
\end{description}
\item[08/E] Drawing, architectural restoration and history
\begin{description}
\item[08/E1] Drawing
\item[08/E2] Architectural restoration and history
\end{description}
\item[08/F] Urban and landscape planning and design
\begin{description}
\item[08/F1] Urban and landscape planning and design
\end{description}
\end{description}
\item[09] Industrial and information engineering
\begin{description}
\item[09/A] Mechanical and aerospace engineering and naval architecture
\begin{description}
\item[09/A1] Aeronautical and aerospace engineering and naval architecture
\item[09/A2] Applied mechanics
\item[09/A3] Industrial design, machine construction and metallurgy
\end{description}
\item[09/B] Manufacturing, industrial and managenent engineering
\begin{description}
\item[09/B1] Manufacturing technology and systems
\item[09/B2] Industrial mechanical plants
\item[09/B3] Business and management engineering
\end{description}
\item[09/C] Energy, thermomechanical and nuclear engineering
\begin{description}
\item[09/C1] Fluid machinery, energy systems and power generation
\item[09/C2] Technical physics and nuclear engineering
\end{description}
\item[09/D] Chemical and materials engineering
\begin{description}
\item[09/D1] Materials science and technology
\item[09/D2] Systems, methods and technologies of chemical and process engineering
\item[09/D3] Chemical plants and technologies
\end{description}
\item[09/E] Electrical and electronic engineering and measurements
\begin{description}
\item[09/E1] Electrical technology
\item[09/E2] Electrical energy engineering
\item[09/E3] Electronics
\item[09/E4] Measurements
\end{description}
\item[09/F] Telecommunications engineering and electromagnetic fields
\begin{description}
\item[09/F1] Electromagnetic fields
\item[09/F2] Telecommunications
\end{description}
\item[09/G] Systems engineering and bioengineering
\begin{description}
\item[09/G1] Systems and control engineering
\item[09/G2] Bioengineering
\end{description}
\item[09/H] Computer engineering
\begin{description}
\item[09/H1] Information processing systems
\end{description}
\end{description}
\item[10] Antiquities, philology, literary studies, art history
\begin{description}
\item[10/A] Archaeological sciences
\begin{description}
\item[10/A1] Archaeology
\end{description}
\item[10/B] Art history
\begin{description}
\item[10/B1] Art history
\end{description}
\item[10/C] Cinema, music, performing arts, television and media studies
\begin{description}
\item[10/C1] Cinema, music, performing arts, television and media studies
\end{description}
\item[10/D] Sciences of antiquity
\begin{description}
\item[10/D1] Ancient history
\item[10/D2] Greek language and literature
\item[10/D3] Latin language and literature
\item[10/D4] Classical and late antique philology
\end{description}
\item[10/E] Medieval latin and romance philologies and literatures
\begin{description}
\item[10/E1] Medieval latin and romance philologies and literatures
\end{description}
\item[10/F] Italian studies and comparative literatures
\begin{description}
\item[10/F1] Italian literature, literary criticism and comparative literature
\item[10/F2] Contemporary Italian literature
\item[10/F3] Italian linguistics and philology
\end{description}
\item[10/G] Glottology and linguistics
\begin{description}
\item[10/G1] Glottology and linguistics
\end{description}
\item[10/H] French studies
\begin{description}
\item[10/H1] French language, literature and culture
\end{description}
\item[10/I] Spanish and Hispanic studies
\begin{description}
\item[10/I1] Spanish and Hispanic languages, literatures and cultures
\end{description}
\item[10/L] English and Anglo-American studies
\begin{description}
\item[10/L1] English and Anglo-American languages, literatures and cultures
\end{description}
\item[10/M] Germanic and Slavic languages, literatures and cultures
\begin{description}
\item[10/M1] Germanic languages, literatures and cultures
\item[10/M2] Slavic studies
\end{description}
\item[10/N] Eastern cultures
\begin{description}
\item[10/N1] Ancient Near Eastern, Middle Eastern and African cultures
\item[10/N3] Central and East Asian cultures
\end{description}
\end{description}
\item[11] History, philosophy, pedagogy and psychology
\begin{description}
\item[11/A] History
\begin{description}
\item[11/A1] Medieval history
\item[11/A2] Modern history
\item[11/A3] Contemporary history
\item[11/A4] Science of books and documents, history of religions
\item[11/A5] Demography, ethnography and anthropology
\end{description}
\item[11/B] Geography
\begin{description}
\item[11/B1] Geography
\end{description}
\item[11/C] Philosophy
\begin{description}
\item[11/C1] Theoretical philosophy
\item[11/C2] Logic, history and philosophy of science
\item[11/C3] Moral philosophy
\item[11/C4] Aesthetics and philosophy of languages
\item[11/C5] History of philosophy
\end{description}
\item[11/D] Educational theories
\begin{description}
\item[11/D1] Educational theories and history of educational theories
\item[11/D2] Methodologies of teaching, special education and educational research
\end{description}
\item[11/E] Psychology
\begin{description}
\item[11/E1] General psychology, psychobiology and psychometrics
\item[11/E2] Developmental and educational psychology
\item[11/E3] Social psychology and work and organizational psychology
\item[11/E4] Clinical and dynamic psychology
\end{description}
\end{description}
\item[12] Law studies
\begin{description}
\item[12/A] Private law
\begin{description}
\item[12/A1] Private law
\end{description}
\item[12/B] Business, navigation and air law and labour law
\begin{description}
\item[12/B1] Business, navigation and air law
\item[12/B2] Labour law
\end{description}
\item[12/C] Constitutional and ecclesiastical law
\begin{description}
\item[12/C1] Constitutional law
\item[12/C2] Ecclesiastical law and canon law
\end{description}
\item[12/D] Administrative and tax law
\begin{description}
\item[12/D1] Administrative law
\item[12/D2] Tax law
\end{description}
\item[12/E] International and European Union law, comparative, economics and markets law
\begin{description}
\item[12/E1] International and European Union law
\item[12/E2] Comparative law
\item[12/E3] Economics, financial and agri-food markets law and regulation
\end{description}
\item[12/F] Civil procedural law
\begin{description}
\item[12/F1] Civil procedural law
\end{description}
\item[12/G] Criminal law and criminal procedure
\begin{description}
\item[12/G1] Criminal law
\item[12/G2] Criminal procedure
\end{description}
\item[12/H] Roman law, history of medieval and modern law and philosophy of law
\begin{description}
\item[12/H1] Roman and ancient law
\item[12/H2] History of medieval and modern law
\item[12/H3] Philosophy of law
\end{description}
\end{description}
\item[13] Economics and statistics
\begin{description}
\item[13/A] Economics
\begin{description}
\item[13/A1] Economics
\item[13/A2] Economic policy
\item[13/A3] Public economics
\item[13/A4] Applied economics
\item[13/A5] Econometrics
\end{description}
\item[13/B] Business administration and Management
\begin{description}
\item[13/B1] Business administration and Management
\item[13/B2] Management
\item[13/B3] Organization studies
\item[13/B4] Financial Markets and Institutions
\item[13/B5] Commodity science
\end{description}
\item[13/C] Economic history
\begin{description}
\item[13/C1] Economic history
\end{description}
\item[13/D] Statistics and mathematical methods for decisions
\begin{description}
\item[13/D1] Statistics
\item[13/D2] Economic statistics
\item[13/D3] Demography and social statistics
\item[13/D4] Mathematical methods of economics, finance and actuarial sciences
\end{description}
\end{description}
\item[14] Political and social sciences
\begin{description}
\item[14/A] Political theory
\begin{description}
\item[14/A1] Political philosophy
\item[14/A2] Political science
\end{description}
\item[14/B] Political history
\begin{description}
\item[14/B1] History of political thought and institutions
\item[14/B2] History of international relations and of non-European societies and institutions
\end{description}
\item[14/C] Sociology
\begin{description}
\item[14/C1] General and political sociology, sociology of law
\item[14/C2] Sociology of culture and communication
\end{description}
\item[14/D] Applied sociology
\begin{description}
\item[14/D1] Sociology of economy and labour, sociology of land and environment\end{description}
\end{description}
\end{description}\end{scriptsize}
\end{multicols}

\section{Descriptive Statistics}\label{app:statistics}

In this section we report some descriptive statistics that can be
derived from the application forms. The statistics provide useful
contextual information on the demography and behavior of applicants,
including: the age distribution, the frequency of publication types
and scientific titles in each area, and the structure of the
co-qualification graph.

\paragraph{Age distribution of applicants}

\begin{figure}[tT]
\centering%
\includegraphics[scale=\figscale]{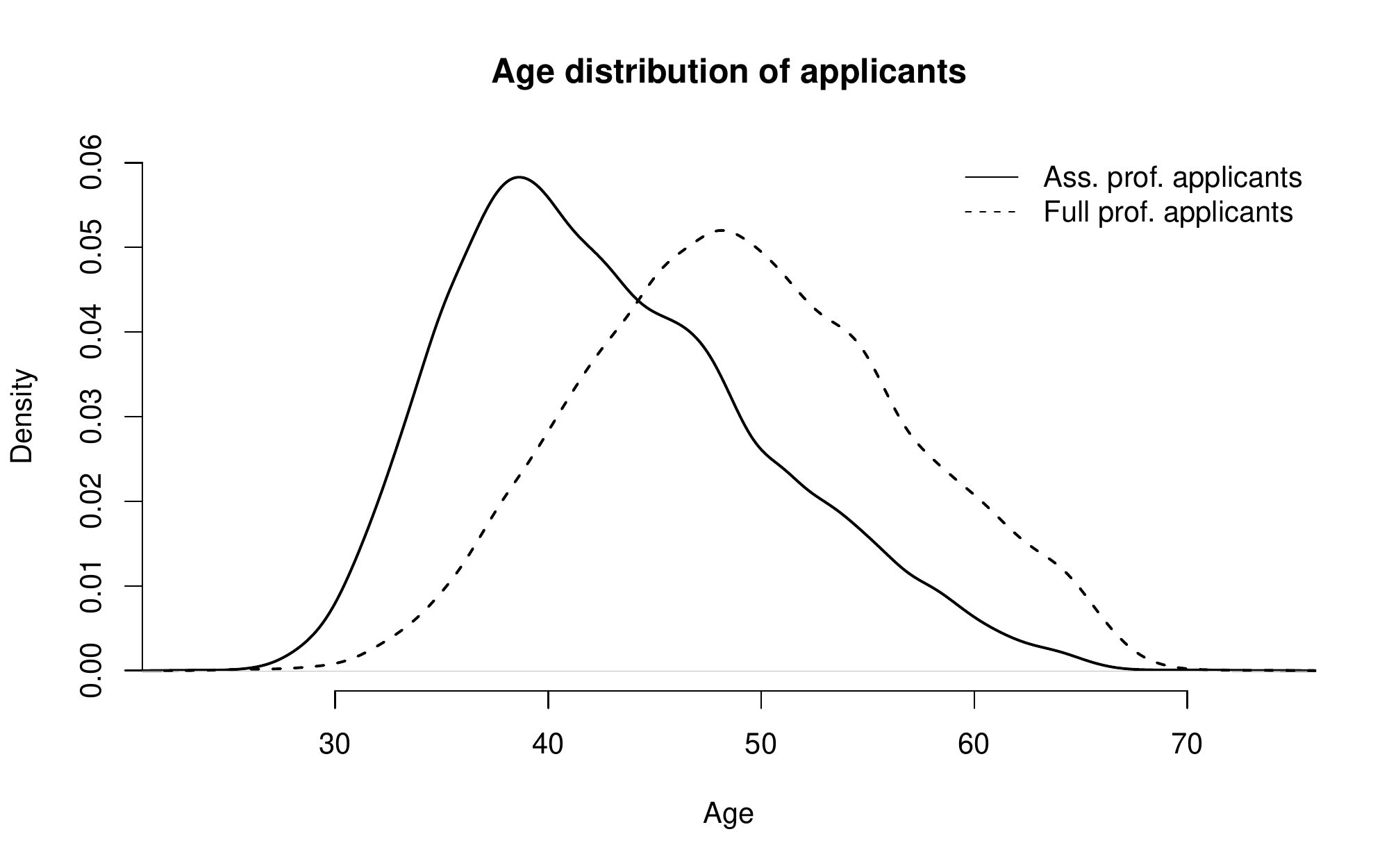}\\
  \begin{small}%
    \begin{fivenum}
      Associate professor applicants& 23 & 37 & 42 & 48 & 74\\
      Full professor applicants& 25 & 44 & 49 & 54 & 74\\
    \end{fivenum}%
  \end{small}
\caption{Age distribution of applicants.}\label{fig:age-distr-all}
\end{figure}

Figure~\ref{fig:age-distr-all} shows the age distribution of
applicants for the full and associate role; individuals applying for
multiple qualifications are counted once per role. The five number
summary shows that applicants for the full professor role are, on
average, slightly older than those applying for the associate level:
the sample median is $49$ years
for full and $42$ years for the
associate role.

\begin{figure}[tT]
\centering\includegraphics[width=.9\textwidth]{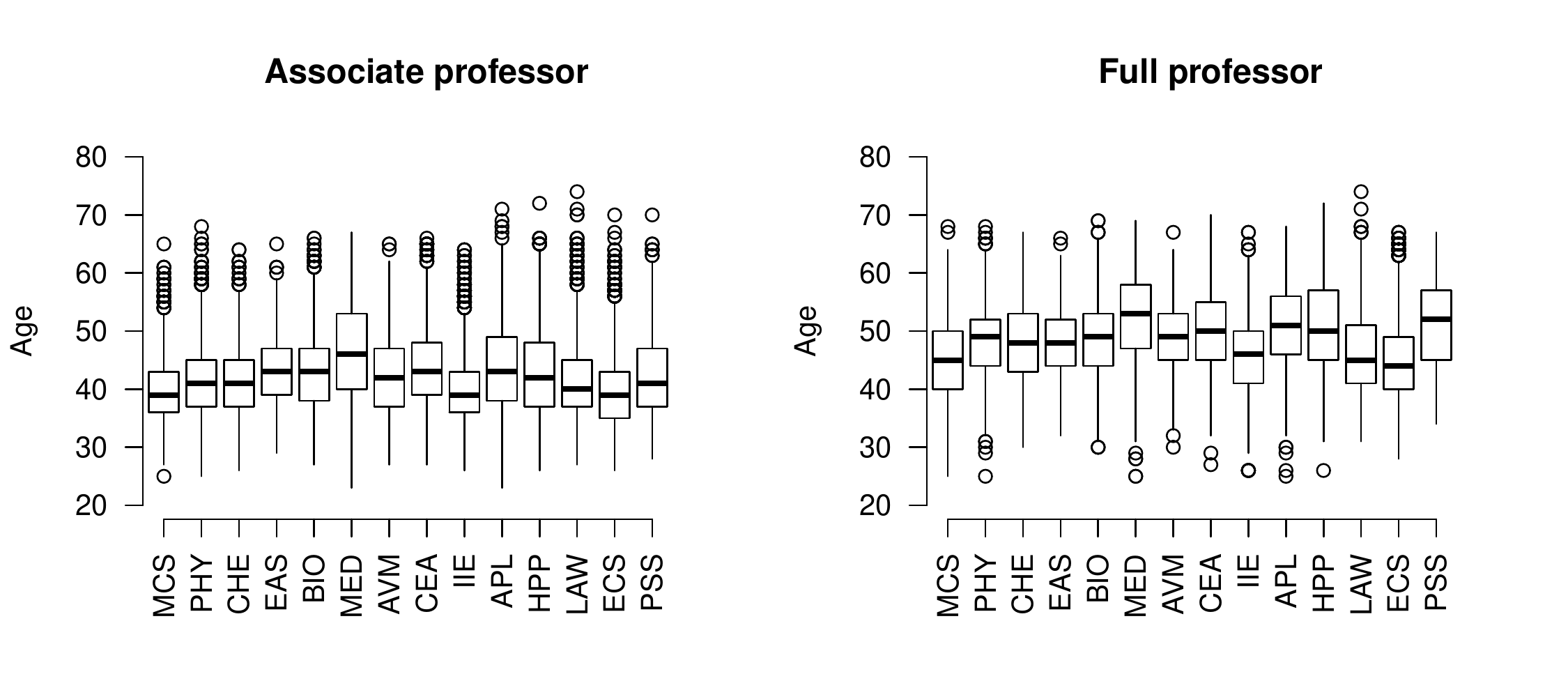}
\caption{Age distribution of applicants by area}\label{fig:age-distr}
\end{figure}

Looking at the individual scientific areas
(Figure~\ref{fig:age-distr}) we observe that the age of applicants
spans a large range. Area~\acf{MED} has the highest median age for
both associate
($46$ years) and
full professor applicants
($53$ years).
The youngest successful applicant
was~$27$ years old (in 2012),
while the oldest was~$69$ years
old. It is worth noticing that the retirement age for university
professors in Italy is currently set to 70 years; yet,
$12$ qualified applicants for the associate
and~$85$ for the full professor role are over 65
years old. These applicants are unlikely to be promoted before they
retire.

Are older applicants more (less) likely to get qualification than
younger ones? To answer this question we use a probit regression
model~\citep{Bliss1934} to study the dependency of the~\ac{ASN} result
(qualified/not qualified) on the applicant's age. A probit model
assumes that the qualification probability for a given age $x$ can be
expressed as:
\begin{equation}
\mathrm{Pr}(\mathit{Qualified}\ |\ \mathit{Age} = x ) = \Phi( \beta \times x )\label{eq:probit}
\end{equation}
\noindent for a suitable scalar parameter $\beta$, where $\Phi(\cdot)$
is the~cumulative distribution function of the normal
distribution. Positive values of $\beta$ denote that older applicants
are more likely to qualify, while negative values denote negative
correlation.

\begin{table}[tT]
  \centering%
  \begin{tabular*}{\textwidth}{@{\extracolsep{\fill}}llrcrc}
    \toprule
    & {\em Area} & \multicolumn{4}{c}{95\% CI for $\beta$} \\
    & & \multicolumn{2}{l}{Full professor} & \multicolumn{2}{l}{Associate professor} \\
    \midrule
MCS&\acl{MCS} & $[-0.0270, -0.0002]$& -& $[-0.0461, -0.0225]$& -\\
PHY&\acl{PHY} & $[-0.0169, 0.0025]$&& $[0.0015, 0.0157]$& +\\
CHE&\acl{CHE} & $[-0.0334, -0.0055]$& -& $[-0.0211, -0.0003]$& -\\
EAS&\acl{EAS} & $[-0.0070, 0.0360]$&& $[0.0098, 0.0382]$& +\\
BIO&\acl{BIO} & $[-0.0037, 0.0138]$&& $[0.0011, 0.0122]$& +\\
MED&\acl{MED} & $[-0.0106, 0.0014]$&& $[-0.0142, -0.0062]$& -\\
AVM&\acl{AVM} & $[-0.0196, 0.0156]$&& $[-0.0030, 0.0170]$&\\
CEA&\acl{CEA} & $[0.0002, 0.0224]$& +& $[-0.0095, 0.0058]$&\\
IIE&\acl{IIE} & $[-0.0053, 0.0162]$&& $[-0.0237, -0.0066]$& -\\
APL&\acl{APL} & $[0.0028, 0.0187]$& +& $[0.0021, 0.0119]$& +\\
HPP&\acl{HPP} & $[-0.0018, 0.0177]$&& $[-0.0290, -0.0166]$& -\\
LAW&\acl{LAW} & $[-0.0444, -0.0192]$& -& $[-0.0562, -0.0367]$& -\\
ECS&\acl{ECS} & $[-0.0403, -0.0228]$& -& $[-0.0393, -0.0242]$& -\\
PSS&\acl{PSS} & $[-0.0060, 0.0281]$&& $[-0.0194, 0.0034]$&\\\bottomrule
\end{tabular*}
\caption{Confidence intervals for $\beta$ (Eq.~\ref{eq:probit}). '+'
  denotes positive correlation between age and qualification
  probability, i.e., older applicants are more likely to qualify; '-'
  denotes negative correlation.}\label{tab:z-value}
\end{table}

Table~\ref{tab:z-value} shows 95\%~\acp{CI} for the value of $\beta$
for each area and role. Positive correlation is observed, among
others, for both roles in areas~\ac{MCS}, \ac{CHE}, \ac{LAW}
and~\ac{ECS}. Negative correlation is observed in area~\ac{APL}. Where
the~\ac{CI} for $\beta$ includes zero, we can not reject the
hypothesis that the qualification probability is unrelated to the age.

\paragraph{Distribution of publication types}

\begin{table}[tT]
  \centering\begin{small}%
  \begin{tabular*}{\textwidth}{@{\extracolsep{\fill}}lrrr}
    \toprule
    {\em Publication Type} & {\em Count} & $\%$ & {\em Rank} \\
    \midrule
{\bf Journal contribution} & $\mathbf{2276633}$ & $\mathbf{62.68}$\\
 Journal paper & 2115083 & 58.23 & 1 \\ 
  Abstract in journal & 100142 & 2.76 & 5 \\ 
  Review in journal & 49099 & 1.35 & 8 \\ 
  Comment of verdict & 9457 & 0.26 & 14 \\ 
  Translation in journal & 2402 & 0.07 & 21 \\ 
  Bibliography & 450 & 0.01 & 33 \\ {\bf Volume contribution} & $\mathbf{417025}$ & $\mathbf{11.48}$\\
 Book chapter & 356326 & 9.81 & 3 \\ 
  Dictionary or encyclopedia entry & 28635 & 0.79 & 10 \\ 
  Catalogue entry & 15476 & 0.43 & 12 \\ 
  Preface/postface & 7881 & 0.22 & 15 \\ 
  Translation in volume & 4382 & 0.12 & 17 \\ 
  Introduction & 3529 & 0.10 & 18 \\ 
  Review in volume & 796 & 0.02 & 28 \\ {\bf Book} & $\mathbf{93475}$ & $\mathbf{2.57}$\\
 Monograph or scientific treatise & 80800 & 2.22 & 6 \\ 
  Book translation & 4935 & 0.14 & 16 \\ 
  Bibliographic entry & 3209 & 0.09 & 19 \\ 
  Critical edition of books/archaeological excavation & 2676 & 0.07 & 20 \\ 
  Scientific commentary & 791 & 0.02 & 29 \\ 
  Publication of new literary or archivistic document & 647 & 0.02 & 31 \\ 
  Index & 260 & 0.01 & 36 \\ 
  Concordance & 157 & 0.00 & 37 \\ {\bf Contribution in proceedings} & $\mathbf{728415}$ & $\mathbf{20.05}$\\
 Paper in proceedings & 538856 & 14.84 & 2 \\ 
  Abstract in proceedings & 164951 & 4.54 & 4 \\ 
  Poster & 24608 & 0.68 & 11 \\ {\bf Patents} & $\mathbf{14446}$ & $\mathbf{0.4}$\\
 Patent & 14446 & 0.40 & 13 \\ {\bf Curatorship} & $\mathbf{40196}$ & $\mathbf{1.11}$\\
 Curatorship & 40196 & 1.11 & 9 \\ {\bf Other} & $\mathbf{62064}$ & $\mathbf{1.71}$\\
 Other publication types & 50554 & 1.39 & 7 \\ 
  Composition & 2043 & 0.06 & 22 \\ 
  Database & 1732 & 0.05 & 23 \\ 
  Exhibition & 1604 & 0.04 & 24 \\ 
  Software & 1497 & 0.04 & 25 \\ 
  Exposition & 1324 & 0.04 & 26 \\ 
  Chart & 1133 & 0.03 & 27 \\ 
  Drawing & 660 & 0.02 & 30 \\ 
  Design & 591 & 0.02 & 32 \\ 
  Performance & 401 & 0.01 & 34 \\ 
  Artifact & 373 & 0.01 & 35 \\ 
  Art prototype & 152 & 0.00 & 38 \\ \midrule
{\bf Total} & $3632254$ & $100.00$ \\
\bottomrule
  \end{tabular*}\end{small}
\caption{Counts of publication types. Percentages refer to the
  fraction of each type with respect to the total number of
  publications submitted by all applicants; \emph{Rank} is the rank of
  each type according to the frequency of occurrence in the
  CVs.}\label{tab:publication-types}
\end{table}

The publications that can be listed in the applications forms are
divided into seven categories: journal contribution, volume
contribution, book, contribution in proceeding, patent, curatorship,
and other publication type. Table~\ref{tab:publication-types} shows
the list of the seven main categories and all sub-categories with
their counts. The same publication may be counted multiple times,
e.g., if it has multiple authors that are applying for qualification,
or one of the authors applied for qualification on several disciplines
or roles. We did not attempt to remove duplicates, since that would
have had little impact on the rank of publication types at the cost of
considerable technical complexity.

The five most frequent types -- journal article, paper in proceedings,
book chapter, abstract in proceedings, and abstract in journal,
respectively -- represent more than 90\% of all publications appearing
in the dataset. The small but non-negligible fraction of ``Other
publication types'' (1.39\%) consists mostly of technical reports that
have not been formally published.

Each~\ac{SD} has its own practices regarding the preferred venues for
disseminating their research output; these differences are apparent if
we look at Table~\ref{tab:top-pubtypes} in the Appendix, that lists
the four most common publication types for each~\ac{SD}. Journal
papers are common in areas~\acl{MCS}, \acl{PHY}, \acl{CHE}, \acl{EAS},
\acl{BIO}, \acl{MED}, and~\acl{AVM}, with the notable exception of
01/B1--\emph{Computer Science} where the most common publication type
is the conference proceeding.  This peculiarity of 01/B1 is in
accordance with the DBLP computer science bibliography, that indexes
2.6 million publications by 1.4 million authors; at the time of
writing, 55.99\% of the bibliographic entries in DBLP are conference
proceedings, and 39.94\% are journal
papers\footnote{\url{http://dblp.uni-trier.de/statistics/distributionofpublicationtype},
  accessed on 2015-10-06.}.

A common trait of the areas above, apart from a few cases, is that the
four most common publication types account for more than 90\% of the
total number of publications. In the remaining areas (\acl{CEA},
\acl{IIE}, \acl{APL}, \acl{HPP}, \acl{LAW}, \acl{ECS}, and~\acl{PSS}),
the most frequent publication type is again the journal article, with
a significant number of disciplines where conference proceedings or
book chapters are the preferred media. Interestingly, the social
sciences and humanities adopt more diversified dissemination
practices: the four most frequent publication types account for about
70\%--80\% of the publications.

While there are yet no comprehensive studies on the frequency of
publication types on different scientific areas, some data have been
analyzed for Norway and Australia. \cite{sivertsen09} analyzes the
frequency of articles in journals (with ISSN), articles in books (with
ISBN), and books for all scientific fields in Norway higher education
sector; articles in books here include also papers in conference
proceedings. The data shows that publication patterns are quite
different across~\acp{SD} and also within subfields of the same
discipline, in particular within the social sciences and humanities.
This is in accordance with our findings
(see~\ref{app:pubtypes}). Also, publication types in the computer
science community in Norway show the same skewness towards conference
papers that we observe.

The report of the~\ac{ERA} evaluation~\citep{era} contains statistics
on the publications submitted as part of the national evaluation of
Australian universities and research institutes.  Caution should be
adopted in comparing~\ac{ERA} and~\ac{ASN}, since they have very
different goals -- \ac{ERA} aims at evaluating research institutes,
while the~\ac{ASN} evaluates individuals.
  
\ac{ERA} classifies research outputs in three main categories:
  
\begin{description}
\item[Traditional outputs:] Books, book chapters, conference
  publications and journal articles;
\item[Non-traditional outputs:] Curated or exhibited event, live
  performance, original creative work, recorded/rendered work,
  portfolio of non-traditional research outputs;
\item[Output types within portfolios:] Curated exhibited events,
  live performance, original creative work, recorded rendered
  work.
\end{description}

More than $413,000$ research outputs were submitted to the~\ac{ERA}:
69\% were journal articles, 18\% conference papers, 10\% book
chapters, 1\% books, and the remaining 2\% non-traditional
outputs. These percentages are remarkably similar to the percentages
of journal contributions, contributions in proceedings, volume
contributions and books shown in
Table~\ref{tab:publication-types}. Looking at individual disciplines,
62\% of research outputs within the~\ac{ERA} research area
``Information and computing sciences'' are conference papers, 30\% are
journal articles, 7\% book chapters, and less than 1\% books. These
are similar to those observed in our dataset for 01/B1--\emph{Computer
  Science}.
  
Table~\ref{tab:top-pubtypes} shows that abstracts are unusually common
in many~\ac{ASN} disciplines, in particular those of areas~5
(\ac{BIO}) and~6 (\ac{MED}). For example, abstracts represent more
than 20\% of all publications listed in the curricula of applicants
for~06/E2--\emph{Plastic and paediatric surgery and urology}.  Since
the rank of publication types remains the same even if we consider
successful applicants only, abstracts are not used by low quality
applicants only, but instead play an important role in the
dissemination of research results in some scientific communities.
  
The role of abstracts that emerges from our dataset is more prominent
than what can be desumed from other sources. For example, while
abstracts represent 15\% of the publications of successful
qualifications in area~\ac{MED}, they constitute only 4\% of the
references listed by PubMed, a bibliographic database of biomedical
research
papers\footnote{\url{https://www.nlm.nih.gov/bsd/licensee/2014_stats/2014_less_OLDMEDLINE_LO.html},
  accessed on 2015-10-06.}. The origin of this difference should be
investigated in future studies.

\paragraph{Distribution of scientific titles}

\begin{table}[tT]
  \centering\begin{small}%
  \begin{tabular*}{\textwidth}{@{\extracolsep{\fill}}p{.42\textwidth}rrrrrr}
    \toprule
        {\em Scientific Title} & \multicolumn{3}{c}{\em Associate} & \multicolumn{3}{c}{\em Full} \\
        \cmidrule{2-4}\cmidrule{5-7}
        & {\em Count} & {\em $\%$ Appl.} & {\em Rank} & {\em Count} & {\em $\%$ Appl.} & {\em Rank} \\
    \midrule
 Other titles & 28459 & 76.60 & 1 & 12936 & 77.69 & 1 \\ 
  Participation to research projects & 27754 & 74.70 & 2 & 1 & 0.01 & 10 \\ 
  Research or teaching fellowships abroad & 18192 & 48.96 & 3 & 9246 & 55.53 & 3 \\ 
  Scientific awards & 16566 & 44.59 & 4 & 8135 & 48.86 & 5 \\ 
  Membership of editorial board of journals & 13954 & 37.56 & 5 & 8837 & 53.07 & 4 \\ 
  Involvement with foreign research institutes & 11521 & 31.01 & 6 & 1 & 0.01 & 10 \\ 
  Technology transfer activities (e.g., startups) & 5548 & 14.93 & 7 & 3642 & 21.87 & 7 \\ 
  Direction of research institutes & 1 & 0.00 & 9 & 1466 & 8.80 & 9 \\ 
  Membership of scientific academies & 1 & 0.00 & 9 & 3661 & 21.99 & 6 \\ 
  Coordination of research projects & 1 & 0.00 & 9 & 11275 & 67.71 & 2 \\ 
  Editor in chief of journals, encyclopedias, or treatises & 0 & 0.00 & 11 & 2796 & 16.79 & 8 \\ \midrule
{\bf Number of applications} & $37154$ &&& $16651$ &&\\
    \bottomrule
  \end{tabular*}\end{small}
\caption{Application counts with at least one instance of a given
  scientific title. Percentages refer to the fraction of applications
  with at least one instance of the given title, therefore the
  percentages do not sum to 100.}\label{tab:titles}
\end{table}

The last part of the application forms contain the list of additional
scientific qualifications (also called \emph{scientific titles}) of
the candidate. The list of allowed scientific titles, that is the same
for both associate and full professor applicants, is reported in
Table~\ref{tab:titles}. Candidates were required to supply additional
details in some cases; for example, an applicant claiming
``Participation to research projects'' had to specify the project
name, duration and role assumed (e.g., participant, task coordinator,
affiliate member).

The most frequently mentioned title, appearing in
$76.6\%$ of the applications for the
associate and $77.69\%$ for the full
professor role, is the catch-all category ``Other titles''. Manual
examination reveals that candidates used this category to list
teaching duties, service activities (conference organization,
coordination of Master or PhD programs, program committee
memberships), invited presentations and consulting activities. All
these items seems relevant, and the fact that they appear frequently
suggests that they should be given specific entries on their own.  

Teaching experience is a conspicuous omission from the list of
qualifications; research and teaching fellowships at \emph{foreign}
universities can be indicated, but teaching activities at Italian
institutions can not. While the~\ac{ASN} is intended to attest only
the scientific qualification of applicants, professors at Italian
universities are required to teach (there are no research-only
positions in Italy).

Table~\ref{tab:titles} shows a couple of differences between associate
and full professor applications. ``Coordination of research
projects'', ``Editor in chief of journals, encyclopedias, or
treatises'', ``Membership of scientific academies'' and ``Direction of
research institutes'' are claimed by applicants for the full professor
role only, with a single exception. This is understandable, since
these roles, in particular direction of research institutes, are
usually held by well-established scientists that are likely
approaching the top of the academic rank. Note that department heads
and team leaders of Italian national research centers (CNR, INFN,
ENEA...), are not necessarily university professors, and some of them
applied to the~\ac{ASN} claiming (correctly) direction of research
institutes. Interestingly, of the $14,67$ applications claiming
direction of research institutes, only 762 were successful.

On the other hand, ``Participation to research projects'' and
``Involvement with research institutes'' are claimed by candidates for
associate professor qualification only, again with a single
exception. We see no obvious reason why applicants for the higher role
should not pursue these activities; perhaps they are just considered
not worth being mentioned.

\paragraph{Co-qualification analysis}

\begin{table}[tT]
  \centering%
  \begin{tabular*}{\textwidth}{@{\extracolsep{\fill}}lrrrr}
    \toprule
    $n$ & {\em Number of applicants that}      & $\%$ & {\em Number of qualified applicants that}        & $\%$ \\
        & {\em submitted $n$ applications} &      & {\em acquired $n$ qualifications}  &      \\
    \midrule
1 & 27374 & $73.37$ & 17123 & $86.50$\\
2 & 6726 & $18.03$ & 2071 & $10.46$\\
3 & 1670 & $4.48$ & 397 & $2.01$\\
4 & 853 & $2.29$ & 136 & $0.69$\\
5 & 259 & $0.69$ & 35 & $0.18$\\$>5$ & 430 & $1.15$ & 34 & $0.17$\\\midrule
Total & 37312 & $100.00$ & 19796 & $100.00$\\    \bottomrule
  \end{tabular*}
\caption{Number of individuals that submitted $n$ applications; number
  of applicants that received $n$ qualifications.}\label{tab:counts}
\end{table}

The~\ac{ASN} allowed individuals to apply for qualification in
multiple~\acp{SD} and roles. Table~\ref{tab:counts} shows how many
candidates submitted $n$ different applications, and how many received
$n$ qualifications. Our dataset contains~$53,805$
applications from~$37,312$ individuals.  Most of
the applicants ($73.37\%$) submitted a single application, but a
significant fraction ($18.03\%$) submitted two. The maximum number of
applications submitted by one individual
is~$34$ (none of them was
successful). Overall, $19,796$
applicants were granted at least one qualification; $86.50\%$ of them
acquired exactly one qualification, and $10.46\%$ got two. The most
successful applicant qualified for both roles in~8~\acp{SD},
collecting a total of~16 qualifications.

  \begin{table}[tT]
    \centering%
    \begin{fivenum}
      & 0.001 & 0.004 & 0.006 & 0.014 & 0.338\\
    \end{fivenum}
    \caption{Five number summary of the nonzero entries of the
      co-qualification matrix.}\label{tab:co-strength}
  \end{table}

The existence of individuals that qualified in two different~\ac{SD},
say $i$ and $j$, is an indication that some overlap may exist between
the scope of $i$ and $j$, fostered by the personal interest of
researchers working on cross-disciplinary boundaries. In this section
we study \emph{co-qualifications} in more detail, as a proxy for the
level of affinity among~\acp{SD}.

For each pair of disciplines $i, j$, $i \neq j$, we define the
\emph{co-qualification strength} $M_{ij}$ as the fraction of
applicants that qualified in either $i$ or $j$ that qualified in both:
\[
M_{ij} = \frac{\mbox{N. of applicants that qualified in \emph{both} SD $i$ and $j$}}{\mbox{N. of applicants that qualified in \emph{either} SD $i$ or $j$}}
\]
By definition, $0 \leq M_{ij} \leq 1$ and $M_{ij} = M_{ji}$. If
$M_{ij} = 0$, then there is no applicant that received qualification
in both $i$ and $j$; this suggests that disciplines $i$ and $j$ might
be unrelated. $M_{ij} = 1$ means that every applicant that qualified
for~\ac{SD} $i$ also qualified for~$j$. It turns out that
co-qualifications across disciplines are relatively rare: only
$531$ out of $170 \times 169
/ 2 = 14,365$ pairs have nonzero co-qualification
strength; the five number summary of the nonzero values of the
co-qualification matrix are shown in Table~\ref{tab:co-strength}.

\begin{figure}[tT]
  \centering\includegraphics[width=.9\textwidth]{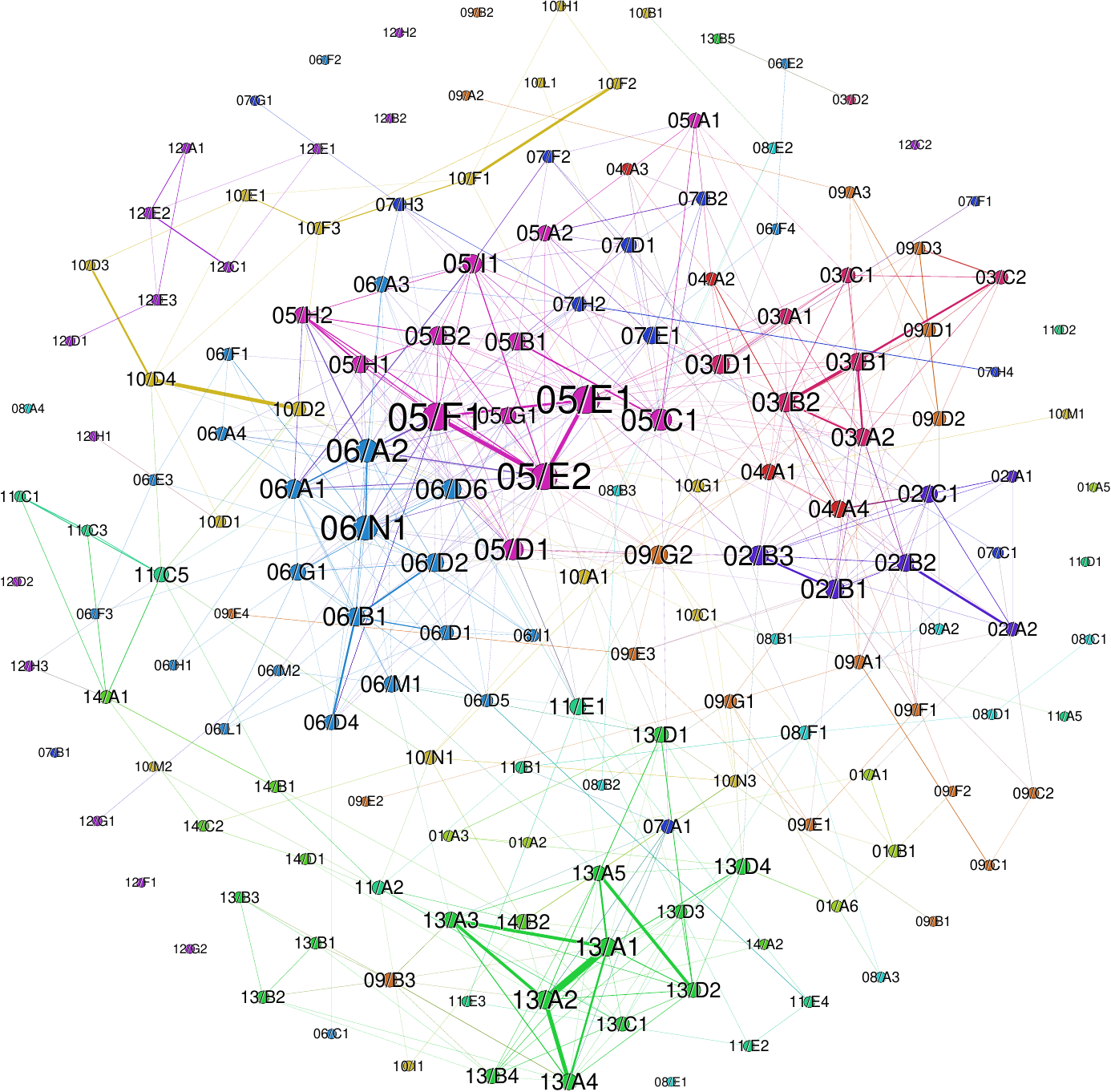}
  \caption{Co-qualification graph (best viewed in color). Colors
    denote the 14 scientific areas. Node sizes are proportional to the
    number of incident edges; edge widths are proportional to
    co-qualification strengths. See text for
    details.}\label{fig:co-graph}
\end{figure}

An effective way to visualize co-qualifications is to draw the
\emph{co-qualification graph} $G$ (Figure~\ref{fig:co-graph}). $G$ is
a weighted, undirected graph where each node represents a~\ac{SD}, and
two nodes $i, j$ are connected by an edge of weight $M_{ij}$ if and
only there exists at least one applicant that qualified in both $i$
and $j$.

The co-qualification graph has~$170$ nodes
and~$531$ edges. We use colors to distinguish the
14 scientific areas. The node sizes are proportional to the number of
incident edges, and edge widths is proportional to the
co-qualification strength: thick edges denote a higher fraction of
co-qualified applicants (i.e., higher values of $M_{ij}$). We used
Gephi~\citep{gephi} and igraph~\citep{igraph} to draw $G$ and compute
the metrics described in the following.

\begin{figure}[tT]
  \centering\includegraphics{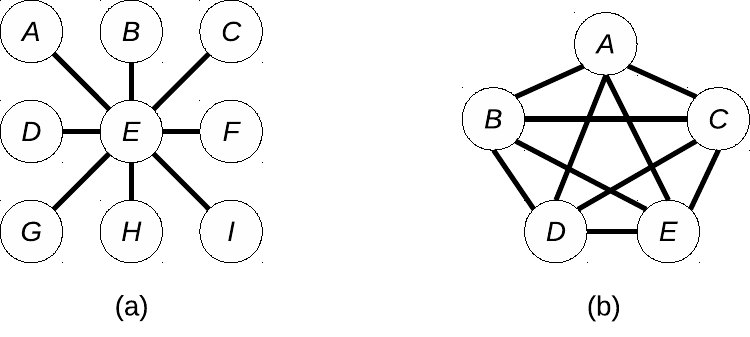}
  \caption{Important sub-structures of the co-qualification graph:
    (a)~hub; (b)~clique}\label{fig:grafi}
\end{figure}

To study the relationships among~\acp{SD} we look for two important
structural patterns in the co-qualification graph: \emph{hubs} and
\emph{cliques}. A hub is a node with a large number of neighbors, such
as node $E$ in Figure~\ref{fig:grafi}~(a). A hub in $G$ can be
interpreted as a ``general'' discipline with partial overlaps with
more specific ones that are not necessarily related each other. A
clique is a complete subgraph, i.e., a subset of nodes that are
pairwise connected by an edge; as an example, nodes $\{A, B, C, D,
E\}$ in Figure~\ref{fig:grafi}~(b) form a clique. Cliques in the
co-qualification graph represent disciplines having mutual overlap,
identifying a broader area of related research activities.

\begin{table}[tT]
  \centering%
  \begin{tabular*}{\textwidth}{@{\extracolsep{\fill}}ll}
    \toprule
    {\em Scientific Discipline} & {\em N. of neighbors} \\
    \midrule
05/F1--\emph{Experimental biology} & 28 \\
05/E1--\emph{General biochemistry and clinical biochemistry} & 28 \\
05/E2--\emph{Molecular biology} & 26 \\
06/N1--\emph{Applied medical technologies} & 23 \\
06/A2--\emph{Experimental medicine, pathophysiology and clinical pathology} & 21 \\
05/C1--\emph{Ecology} & 19 \\
05/D1--\emph{Physiology} & 18 \\
02/B3--\emph{Applied physics} & 16 \\
06/D6--\emph{Neurology} & 16 \\
02/B1--\emph{Experimental physics of matter} & 15 \\\bottomrule
  \end{tabular*}
\caption{The ten disciplines with highest degree in the
  co-qualification graph}\label{tab:top-degree}
\end{table}

Hubs can be identified by looking at the \emph{node degree
  distribution} of $G$. The degree $\delta(v)$ of a node $v$ is the
number of incident edges (an edge is incident to a node if it has one
of the endpoints on that node). The hubs in $G$ are the disciplines
with higher degree.

The ten biggest hubs are shown in Table~\ref{tab:top-degree}. Five of
them (05/F1--\emph{Experimental biology}, 05/E1--\emph{General
  biochemistry and clinical biochemistry}, 05/E2--\emph{Molecular
  biology}, 05/C1--\emph{Ecology}, and 05/D1--\emph{Physiology}) belong
to area~\ac{BIO}; three (06/N1--\emph{Applied medical technologies},
06/A2--\emph{Experimental medicine, pathophysiology and clinical
  pathology}, and 06/D6--\emph{Neurology}) belong to area~\ac{MED},
and the remaining two (02/B3--\emph{Applied physics} and
02/B1--\emph{Experimental physics of matter}) belong to area~\ac{PHY}.

The co-qualification graph contains several cliques, i.e., complete
subgraphs. A \emph{maximal cliques} $G'$ is a subgraph $G' \subseteq
G$ such that no node can be added to $G'$ to form a bigger clique. The
largest clique in $G$ has size 9, and consists of the following
disciplines (all belonging to areas~\ac{BIO} and~\ac{MED}):
\begin{itemize}
  \item
05/H1--\emph{Human anatomy}, 05/F1--\emph{Experimental biology}, 05/H2--\emph{Histology}, 05/B2--\emph{Comparative anatomy and cytology}, 06/N1--\emph{Applied medical technologies}, 06/A1--\emph{Medical genetics}, 06/A2--\emph{Experimental medicine, pathophysiology and clinical pathology}, 05/E2--\emph{Molecular biology}, and 05/E1--\emph{General biochemistry and clinical biochemistry}.\end{itemize}

The ties between disciplines in area~\ac{MED} and~\ac{BIO} are
confirmed by the existence of three maximal cliques of size 8 that
include the following disciplines:
\begin{itemize}
\item 05/I1--\emph{Genetics and microbiology}, 05/F1--\emph{Experimental biology}, 05/B2--\emph{Comparative anatomy and cytology}, 05/H2--\emph{Histology}, 06/A1--\emph{Medical genetics}, 05/E2--\emph{Molecular biology}, 05/E1--\emph{General biochemistry and clinical biochemistry}, and 06/A2--\emph{Experimental medicine, pathophysiology and clinical pathology}.
\item 05/H1--\emph{Human anatomy}, 05/F1--\emph{Experimental biology}, 05/H2--\emph{Histology}, 05/B2--\emph{Comparative anatomy and cytology}, 06/N1--\emph{Applied medical technologies}, 05/D1--\emph{Physiology}, 05/E2--\emph{Molecular biology}, and 05/E1--\emph{General biochemistry and clinical biochemistry}.
\item 05/H1--\emph{Human anatomy}, 05/F1--\emph{Experimental biology}, 05/H2--\emph{Histology}, 05/B2--\emph{Comparative anatomy and cytology}, 06/N1--\emph{Applied medical technologies}, 06/A1--\emph{Medical genetics}, 06/A2--\emph{Experimental medicine, pathophysiology and clinical pathology}, and 06/D6--\emph{Neurology}.\end{itemize}

Other smaller cliques exist:
$5$ maximal cliques
of size 7, $17$
maximal cliques of size 6, and
$133$ maximal cliques
of size between 3 and 5 inclusive.

\section{Most frequent publication types for each scientific discipline}\label{app:pubtypes}

The following table lists the four most frequent publication types for each~\ac{SD}. We use the following keys:
\textbf{ABSJ}~=~Abstract in journal; \textbf{ABSP}~=~Abstract in proceedings; \textbf{AF}~=~Artifact; \textbf{ART}~=~Art prototype; \textbf{BIB}~=~Bibliography; \textbf{BIBE}~=~Bibliographic entry; \textbf{CAT}~=~Catalogue entry; \textbf{CH}~=~Chart; \textbf{CHAP}~=~Book chapter; \textbf{COM}~=~Composition; \textbf{COMM}~=~Scientific commentary; \textbf{CONC}~=~Concordance; \textbf{CRIT}~=~Critical edition of books/archaeological excavation; \textbf{CUR}~=~Curatorship; \textbf{DB}~=~Database; \textbf{DES}~=~Design; \textbf{DICT}~=~Dictionary or encyclopedia entry; \textbf{DRAW}~=~Drawing; \textbf{EXH}~=~Exhibition; \textbf{EXP}~=~Exposition; \textbf{IDX}~=~Index; \textbf{INTRO}~=~Introduction; \textbf{JRNL}~=~Journal paper; \textbf{MONO}~=~Monograph or scientific treatise; \textbf{OP}~=~Other publication types; \textbf{PAT}~=~Patent; \textbf{PERF}~=~Performance; \textbf{POS}~=~Poster; \textbf{PREF}~=~Preface/postface; \textbf{PROC}~=~Paper in proceedings; \textbf{REVJ}~=~Review in journal; \textbf{REVV}~=~Review in volume; \textbf{SRC}~=~Publication of new literary or archivistic document; \textbf{SW}~=~Software; \textbf{TRB}~=~Book translation; \textbf{TRJ}~=~Translation in journal; \textbf{TRV}~=~Translation in volume; \textbf{VERD}~=~Comment of verdict; 
\begin{scriptsize}
  \LTcapwidth=\textwidth
\begin{longtable}{@{\extracolsep{\fill}}llllllllll}
  \caption{Four most frequent publication types for each scientific discipline.\label{tab:top-pubtypes}}\\
  \toprule
  {\em SD} & \multicolumn{8}{l}{\em Most common publication types} & {\em Other}\\
  \cmidrule{2-9}
  & \multicolumn{2}{l}{1st} & \multicolumn{2}{l}{2nd} & \multicolumn{2}{l}{3rd} & \multicolumn{2}{l}{4th} & \\
  \midrule
  \endhead  
01/A1 &JRNL & 1487 $(42.26\%)$ &PROC & 848 $(24.10\%)$ &CHAP & 542 $(15.40\%)$ &DICT & 138 $(3.92\%)$ &504 (14.32\%)\\
01/A2 &JRNL & 3369 $(84.65\%)$ &PROC & 264 $(6.63\%)$ &OP & 144 $(3.62\%)$ &CHAP & 109 $(2.74\%)$ &94 (2.36\%)\\
01/A3 &JRNL & 6564 $(84.38\%)$ &PROC & 567 $(7.29\%)$ &CHAP & 262 $(3.37\%)$ &OP & 237 $(3.05\%)$ &149 (1.91\%)\\
01/A5 &JRNL & 1415 $(65.30\%)$ &PROC & 432 $(19.94\%)$ &CHAP & 172 $(7.94\%)$ &OP & 50 $(2.31\%)$ &98 (4.51\%)\\
01/A6 &JRNL & 1031 $(55.46\%)$ &PROC & 497 $(26.73\%)$ &CHAP & 152 $(8.18\%)$ &ABSP & 89 $(4.79\%)$ &90 (4.84\%)\\
01/B1 &PROC & 13318 $(57.16\%)$ &JRNL & 6353 $(27.26\%)$ &CHAP & 2116 $(9.08\%)$ &CUR & 526 $(2.26\%)$ &988 (4.24\%)\\
02/A1 &JRNL & 157547 $(94.91\%)$ &PROC & 6774 $(4.08\%)$ &OP & 620 $(0.37\%)$ &CHAP & 390 $(0.23\%)$ &672 (0.41\%)\\
02/A2 &JRNL & 23983 $(80.09\%)$ &PROC & 4740 $(15.83\%)$ &CHAP & 517 $(1.73\%)$ &OP & 164 $(0.55\%)$ &542 (1.80\%)\\
02/B1 &JRNL & 41390 $(81.24\%)$ &PROC & 6745 $(13.24\%)$ &CHAP & 1269 $(2.49\%)$ &ABSP & 698 $(1.37\%)$ &848 (1.66\%)\\
02/B2 &JRNL & 20305 $(86.23\%)$ &PROC & 1754 $(7.45\%)$ &CHAP & 791 $(3.36\%)$ &ABSP & 267 $(1.13\%)$ &430 (1.83\%)\\
02/B3 &JRNL & 18162 $(71.45\%)$ &PROC & 4370 $(17.19\%)$ &ABSP & 884 $(3.48\%)$ &CHAP & 859 $(3.38\%)$ &1143 (4.50\%)\\
02/C1 &JRNL & 26605 $(79.91\%)$ &PROC & 4937 $(14.83\%)$ &ABSP & 617 $(1.85\%)$ &CHAP & 447 $(1.34\%)$ &689 (2.07\%)\\
03/A1 &JRNL & 7597 $(68.61\%)$ &PROC & 1617 $(14.60\%)$ &ABSP & 890 $(8.04\%)$ &CHAP & 765 $(6.91\%)$ &204 (1.84\%)\\
03/A2 &JRNL & 16780 $(84.15\%)$ &PROC & 1328 $(6.66\%)$ &CHAP & 732 $(3.67\%)$ &ABSP & 637 $(3.19\%)$ &463 (2.33\%)\\
03/B1 &JRNL & 24713 $(84.11\%)$ &PROC & 1594 $(5.42\%)$ &ABSP & 1415 $(4.82\%)$ &CHAP & 913 $(3.11\%)$ &748 (2.54\%)\\
03/B2 &JRNL & 13866 $(68.40\%)$ &PROC & 3659 $(18.05\%)$ &ABSP & 1174 $(5.79\%)$ &CHAP & 730 $(3.60\%)$ &844 (4.16\%)\\
03/C1 &JRNL & 9291 $(74.31\%)$ &PROC & 1712 $(13.69\%)$ &CHAP & 552 $(4.41\%)$ &ABSP & 434 $(3.47\%)$ &514 (4.12\%)\\
03/C2 &JRNL & 4547 $(61.45\%)$ &PROC & 1585 $(21.42\%)$ &ABSP & 532 $(7.19\%)$ &CHAP & 352 $(4.76\%)$ &383 (5.18\%)\\
03/D1 &JRNL & 10941 $(75.61\%)$ &PROC & 1324 $(9.15\%)$ &ABSP & 948 $(6.55\%)$ &PAT & 437 $(3.02\%)$ &820 (5.67\%)\\
03/D2 &JRNL & 3502 $(52.92\%)$ &PROC & 1648 $(24.90\%)$ &ABSP & 895 $(13.52\%)$ &CHAP & 222 $(3.35\%)$ &351 (5.31\%)\\
04/A1 &JRNL & 6719 $(60.94\%)$ &ABSP & 1509 $(13.69\%)$ &PROC & 1350 $(12.24\%)$ &CHAP & 538 $(4.88\%)$ &910 (8.25\%)\\
04/A2 &JRNL & 6606 $(60.01\%)$ &PROC & 1360 $(12.35\%)$ &ABSP & 977 $(8.88\%)$ &CHAP & 829 $(7.53\%)$ &1236 (11.23\%)\\
04/A3 &JRNL & 3620 $(41.34\%)$ &PROC & 2193 $(25.04\%)$ &CHAP & 1134 $(12.95\%)$ &ABSP & 780 $(8.91\%)$ &1030 (11.76\%)\\
04/A4 &JRNL & 3939 $(65.85\%)$ &PROC & 946 $(15.81\%)$ &ABSP & 481 $(8.04\%)$ &CHAP & 357 $(5.97\%)$ &259 (4.33\%)\\
05/A1 &JRNL & 8641 $(59.94\%)$ &PROC & 2612 $(18.12\%)$ &ABSP & 1201 $(8.33\%)$ &CHAP & 1129 $(7.83\%)$ &832 (5.78\%)\\
05/A2 &JRNL & 1893 $(72.47\%)$ &PROC & 250 $(9.57\%)$ &CHAP & 199 $(7.62\%)$ &ABSP & 154 $(5.90\%)$ &116 (4.44\%)\\
05/B1 &JRNL & 8267 $(63.70\%)$ &ABSP & 1772 $(13.65\%)$ &PROC & 1188 $(9.15\%)$ &CHAP & 1046 $(8.06\%)$ &706 (5.44\%)\\
05/B2 &JRNL & 4347 $(74.51\%)$ &PROC & 628 $(10.76\%)$ &ABSP & 315 $(5.40\%)$ &CHAP & 248 $(4.25\%)$ &296 (5.08\%)\\
05/C1 &JRNL & 9785 $(67.39\%)$ &PROC & 1573 $(10.83\%)$ &ABSP & 1534 $(10.56\%)$ &CHAP & 914 $(6.29\%)$ &714 (4.93\%)\\
05/D1 &JRNL & 8545 $(73.32\%)$ &PROC & 1230 $(10.55\%)$ &ABSJ & 629 $(5.40\%)$ &CHAP & 474 $(4.07\%)$ &777 (6.66\%)\\
05/E1 &JRNL & 31942 $(77.93\%)$ &PROC & 2872 $(7.01\%)$ &ABSP & 1657 $(4.04\%)$ &CHAP & 1476 $(3.60\%)$ &3043 (7.42\%)\\
05/E2 &JRNL & 16448 $(81.51\%)$ &PROC & 1176 $(5.83\%)$ &ABSP & 735 $(3.64\%)$ &ABSJ & 604 $(2.99\%)$ &1215 (6.03\%)\\
05/F1 &JRNL & 26147 $(76.72\%)$ &ABSJ & 2036 $(5.97\%)$ &PROC & 1991 $(5.84\%)$ &ABSP & 1669 $(4.90\%)$ &2239 (6.57\%)\\
05/G1 &JRNL & 11862 $(78.12\%)$ &PROC & 960 $(6.32\%)$ &ABSP & 856 $(5.64\%)$ &ABSJ & 668 $(4.40\%)$ &839 (5.52\%)\\
05/H1 &JRNL & 6237 $(72.72\%)$ &PROC & 1144 $(13.34\%)$ &ABSP & 428 $(4.99\%)$ &ABSJ & 394 $(4.59\%)$ &374 (4.36\%)\\
05/H2 &JRNL & 2966 $(80.80\%)$ &CHAP & 217 $(5.91\%)$ &PROC & 193 $(5.26\%)$ &ABSP & 85 $(2.32\%)$ &210 (5.71\%)\\
05/I1 &JRNL & 3761 $(80.95\%)$ &ABSP & 326 $(7.02\%)$ &CHAP & 250 $(5.38\%)$ &PROC & 155 $(3.34\%)$ &154 (3.31\%)\\
06/A1 &JRNL & 9269 $(73.76\%)$ &ABSJ & 1451 $(11.55\%)$ &ABSP & 784 $(6.24\%)$ &POS & 342 $(2.72\%)$ &720 (5.73\%)\\
06/A2 &JRNL & 22580 $(83.24\%)$ &PROC & 1221 $(4.50\%)$ &CHAP & 992 $(3.66\%)$ &ABSJ & 606 $(2.23\%)$ &1727 (6.37\%)\\
06/A3 &JRNL & 4681 $(68.23\%)$ &PROC & 771 $(11.24\%)$ &ABSP & 625 $(9.11\%)$ &ABSJ & 231 $(3.37\%)$ &553 (8.05\%)\\
06/A4 &JRNL & 10291 $(84.56\%)$ &ABSJ & 635 $(5.22\%)$ &ABSP & 435 $(3.57\%)$ &PROC & 379 $(3.11\%)$ &430 (3.54\%)\\
06/B1 &JRNL & 24506 $(74.67\%)$ &ABSJ & 2827 $(8.61\%)$ &PROC & 2327 $(7.09\%)$ &CHAP & 1367 $(4.17\%)$ &1792 (5.46\%)\\
06/C1 &JRNL & 30621 $(56.47\%)$ &ABSP & 6951 $(12.82\%)$ &PROC & 6589 $(12.15\%)$ &ABSJ & 5830 $(10.75\%)$ &4236 (7.81\%)\\
06/D1 &JRNL & 18763 $(76.95\%)$ &ABSJ & 2044 $(8.38\%)$ &PROC & 1568 $(6.43\%)$ &CHAP & 1060 $(4.35\%)$ &949 (3.89\%)\\
06/D2 &JRNL & 16398 $(76.87\%)$ &ABSP & 1938 $(9.08\%)$ &PROC & 920 $(4.31\%)$ &CHAP & 912 $(4.28\%)$ &1165 (5.46\%)\\
06/D4 &JRNL & 35973 $(76.16\%)$ &ABSJ & 3582 $(7.58\%)$ &ABSP & 2967 $(6.28\%)$ &CHAP & 2375 $(5.03\%)$ &2337 (4.95\%)\\
06/D5 &JRNL & 7876 $(71.24\%)$ &CHAP & 953 $(8.62\%)$ &PROC & 913 $(8.26\%)$ &ABSJ & 573 $(5.18\%)$ &740 (6.70\%)\\
06/D6 &JRNL & 22221 $(77.58\%)$ &ABSJ & 2935 $(10.25\%)$ &CHAP & 1313 $(4.58\%)$ &ABSP & 962 $(3.36\%)$ &1210 (4.23\%)\\
06/E2 &JRNL & 7343 $(51.25\%)$ &ABSP & 2886 $(20.14\%)$ &ABSJ & 1797 $(12.54\%)$ &PROC & 1293 $(9.02\%)$ &1009 (7.05\%)\\
06/E3 &JRNL & 6269 $(64.46\%)$ &ABSP & 1179 $(12.12\%)$ &PROC & 815 $(8.38\%)$ &CHAP & 650 $(6.68\%)$ &812 (8.36\%)\\
06/F1 &JRNL & 11694 $(62.97\%)$ &PROC & 2910 $(15.67\%)$ &ABSP & 1284 $(6.91\%)$ &ABSJ & 1047 $(5.64\%)$ &1635 (8.81\%)\\
06/F2 &JRNL & 2086 $(59.28\%)$ &PROC & 491 $(13.95\%)$ &ABSP & 462 $(13.13\%)$ &ABSJ & 202 $(5.74\%)$ &278 (7.90\%)\\
06/F3 &JRNL & 7486 $(59.48\%)$ &ABSP & 1832 $(14.56\%)$ &PROC & 1500 $(11.92\%)$ &CHAP & 1066 $(8.47\%)$ &702 (5.57\%)\\
06/F4 &JRNL & 6870 $(55.19\%)$ &PROC & 1988 $(15.97\%)$ &ABSP & 1183 $(9.50\%)$ &ABSJ & 979 $(7.87\%)$ &1427 (11.47\%)\\
06/G1 &JRNL & 21611 $(74.44\%)$ &ABSJ & 2601 $(8.96\%)$ &PROC & 1556 $(5.36\%)$ &ABSP & 1397 $(4.81\%)$ &1865 (6.43\%)\\
06/H1 &JRNL & 13354 $(68.32\%)$ &ABSP & 1867 $(9.55\%)$ &PROC & 1408 $(7.20\%)$ &ABSJ & 1386 $(7.09\%)$ &1530 (7.84\%)\\
06/I1 &JRNL & 16560 $(58.21\%)$ &ABSP & 3532 $(12.41\%)$ &ABSJ & 3429 $(12.05\%)$ &PROC & 2130 $(7.49\%)$ &2800 (9.84\%)\\
06/L1 &JRNL & 3154 $(60.42\%)$ &ABSP & 549 $(10.52\%)$ &ABSJ & 459 $(8.79\%)$ &CHAP & 378 $(7.24\%)$ &680 (13.03\%)\\
06/M1 &JRNL & 17965 $(75.45\%)$ &PROC & 2124 $(8.92\%)$ &ABSP & 1344 $(5.64\%)$ &CHAP & 934 $(3.92\%)$ &1442 (6.07\%)\\
06/M2 &JRNL & 6911 $(59.64\%)$ &PROC & 2234 $(19.28\%)$ &ABSP & 1085 $(9.36\%)$ &CHAP & 629 $(5.43\%)$ &728 (6.29\%)\\
06/N1 &JRNL & 17614 $(77.38\%)$ &PROC & 1558 $(6.84\%)$ &ABSJ & 1275 $(5.60\%)$ &ABSP & 1024 $(4.50\%)$ &1291 (5.68\%)\\
07/A1 &JRNL & 2252 $(39.77\%)$ &CHAP & 1623 $(28.66\%)$ &PROC & 1105 $(19.52\%)$ &MONO & 279 $(4.93\%)$ &403 (7.12\%)\\
07/B1 &JRNL & 4043 $(54.81\%)$ &PROC & 2270 $(30.78\%)$ &CHAP & 514 $(6.97\%)$ &ABSP & 251 $(3.40\%)$ &298 (4.04\%)\\
07/B2 &JRNL & 4843 $(57.74\%)$ &PROC & 2022 $(24.11\%)$ &CHAP & 632 $(7.53\%)$ &ABSP & 503 $(6.00\%)$ &388 (4.62\%)\\
07/C1 &JRNL & 1725 $(41.92\%)$ &PROC & 1570 $(38.15\%)$ &CHAP & 377 $(9.16\%)$ &ABSP & 220 $(5.35\%)$ &223 (5.42\%)\\
07/D1 &JRNL & 6609 $(52.01\%)$ &PROC & 2403 $(18.91\%)$ &ABSP & 1801 $(14.17\%)$ &ABSJ & 720 $(5.67\%)$ &1174 (9.24\%)\\
07/E1 &JRNL & 5242 $(51.55\%)$ &PROC & 2202 $(21.65\%)$ &ABSP & 1347 $(13.25\%)$ &CHAP & 704 $(6.92\%)$ &674 (6.63\%)\\
07/F1 &JRNL & 3917 $(56.49\%)$ &PROC & 1740 $(25.09\%)$ &ABSP & 555 $(8.00\%)$ &CHAP & 376 $(5.42\%)$ &346 (5.00\%)\\
07/F2 &JRNL & 2633 $(62.57\%)$ &PROC & 610 $(14.50\%)$ &ABSP & 519 $(12.33\%)$ &CHAP & 303 $(7.20\%)$ &143 (3.40\%)\\
07/G1 &JRNL & 5106 $(51.04\%)$ &PROC & 3077 $(30.76\%)$ &ABSP & 781 $(7.81\%)$ &ABSJ & 371 $(3.71\%)$ &668 (6.68\%)\\
07/H2 &JRNL & 3434 $(55.09\%)$ &PROC & 1777 $(28.50\%)$ &ABSP & 481 $(7.72\%)$ &ABSJ & 244 $(3.91\%)$ &298 (4.78\%)\\
07/H3 &JRNL & 3567 $(62.73\%)$ &PROC & 1374 $(24.16\%)$ &ABSP & 390 $(6.86\%)$ &ABSJ & 126 $(2.22\%)$ &229 (4.03\%)\\
07/H4 &JRNL & 1604 $(50.79\%)$ &PROC & 842 $(26.66\%)$ &ABSP & 347 $(10.99\%)$ &ABSJ & 201 $(6.36\%)$ &164 (5.20\%)\\
08/A2 &PROC & 2787 $(48.95\%)$ &JRNL & 1915 $(33.63\%)$ &CHAP & 651 $(11.43\%)$ &ABSP & 117 $(2.05\%)$ &224 (3.94\%)\\
08/A3 &PROC & 1721 $(40.20\%)$ &JRNL & 1146 $(26.77\%)$ &CHAP & 968 $(22.61\%)$ &MONO & 151 $(3.53\%)$ &295 (6.89\%)\\
08/A4 &PROC & 1606 $(52.67\%)$ &JRNL & 1069 $(35.06\%)$ &CHAP & 260 $(8.53\%)$ &MONO & 32 $(1.05\%)$ &82 (2.69\%)\\
08/B1 &PROC & 1989 $(58.07\%)$ &JRNL & 889 $(25.96\%)$ &CHAP & 260 $(7.59\%)$ &OP & 114 $(3.33\%)$ &173 (5.05\%)\\
08/B2 &PROC & 2272 $(48.76\%)$ &JRNL & 1763 $(37.83\%)$ &CHAP & 278 $(5.97\%)$ &ABSP & 177 $(3.80\%)$ &170 (3.64\%)\\
08/B3 &PROC & 4945 $(61.94\%)$ &JRNL & 1986 $(24.88\%)$ &CHAP & 687 $(8.61\%)$ &OP & 190 $(2.38\%)$ &175 (2.19\%)\\
08/C1 &JRNL & 4268 $(31.82\%)$ &CHAP & 3500 $(26.09\%)$ &PROC & 3018 $(22.50\%)$ &MONO & 793 $(5.91\%)$ &1834 (13.68\%)\\
08/D1 &JRNL & 3975 $(34.79\%)$ &CHAP & 3535 $(30.94\%)$ &MONO & 674 $(5.90\%)$ &CUR & 585 $(5.12\%)$ &2658 (23.25\%)\\
08/E1 &CHAP & 1770 $(34.05\%)$ &PROC & 1377 $(26.49\%)$ &JRNL & 766 $(14.74\%)$ &MONO & 369 $(7.10\%)$ &916 (17.62\%)\\
08/E2 &CHAP & 3575 $(38.04\%)$ &JRNL & 1919 $(20.42\%)$ &PROC & 1277 $(13.59\%)$ &CAT & 630 $(6.70\%)$ &1996 (21.25\%)\\
08/F1 &CHAP & 5447 $(34.94\%)$ &JRNL & 4751 $(30.48\%)$ &PROC & 2082 $(13.36\%)$ &MONO & 866 $(5.56\%)$ &2443 (15.66\%)\\
09/A1 &PROC & 5292 $(56.96\%)$ &JRNL & 3001 $(32.30\%)$ &CHAP & 344 $(3.70\%)$ &OP & 333 $(3.58\%)$ &321 (3.46\%)\\
09/A2 &PROC & 2227 $(57.18\%)$ &JRNL & 1267 $(32.53\%)$ &CHAP & 193 $(4.96\%)$ &OP & 76 $(1.95\%)$ &132 (3.38\%)\\
09/A3 &PROC & 6278 $(53.91\%)$ &JRNL & 4326 $(37.15\%)$ &CHAP & 497 $(4.27\%)$ &ABSP & 147 $(1.26\%)$ &397 (3.41\%)\\
09/B1 &PROC & 2065 $(47.19\%)$ &JRNL & 1840 $(42.05\%)$ &CHAP & 298 $(6.81\%)$ &PAT & 63 $(1.44\%)$ &110 (2.51\%)\\
09/B2 &PROC & 1912 $(53.75\%)$ &JRNL & 1218 $(34.24\%)$ &CHAP & 162 $(4.55\%)$ &MONO & 125 $(3.51\%)$ &140 (3.95\%)\\
09/B3 &PROC & 1771 $(45.79\%)$ &JRNL & 1254 $(32.42\%)$ &CHAP & 578 $(14.94\%)$ &OP & 91 $(2.35\%)$ &174 (4.50\%)\\
09/C1 &PROC & 3857 $(61.53\%)$ &JRNL & 1833 $(29.24\%)$ &CHAP & 281 $(4.48\%)$ &OP & 114 $(1.82\%)$ &183 (2.93\%)\\
09/C2 &PROC & 5338 $(50.77\%)$ &JRNL & 4054 $(38.56\%)$ &CHAP & 395 $(3.76\%)$ &MONO & 256 $(2.43\%)$ &471 (4.48\%)\\
09/D1 &JRNL & 6988 $(55.65\%)$ &PROC & 3906 $(31.11\%)$ &ABSP & 573 $(4.56\%)$ &CHAP & 559 $(4.45\%)$ &531 (4.23\%)\\
09/D2 &JRNL & 3476 $(48.07\%)$ &PROC & 2741 $(37.91\%)$ &CHAP & 421 $(5.82\%)$ &ABSP & 339 $(4.69\%)$ &254 (3.51\%)\\
09/D3 &JRNL & 3464 $(49.51\%)$ &PROC & 2611 $(37.32\%)$ &CHAP & 468 $(6.69\%)$ &ABSP & 234 $(3.34\%)$ &220 (3.14\%)\\
09/E1 &PROC & 2855 $(44.56\%)$ &JRNL & 2739 $(42.75\%)$ &ABSP & 384 $(5.99\%)$ &CHAP & 263 $(4.10\%)$ &166 (2.60\%)\\
09/E2 &PROC & 5841 $(68.73\%)$ &JRNL & 2244 $(26.40\%)$ &CHAP & 159 $(1.87\%)$ &MONO & 83 $(0.98\%)$ &172 (2.02\%)\\
09/E3 &PROC & 6793 $(49.16\%)$ &JRNL & 5939 $(42.98\%)$ &CHAP & 397 $(2.87\%)$ &PAT & 304 $(2.20\%)$ &384 (2.79\%)\\
09/E4 &PROC & 3816 $(61.77\%)$ &JRNL & 1920 $(31.08\%)$ &CHAP & 146 $(2.36\%)$ &ABSP & 111 $(1.80\%)$ &185 (2.99\%)\\
09/F1 &PROC & 4165 $(51.45\%)$ &JRNL & 3359 $(41.49\%)$ &ABSP & 216 $(2.67\%)$ &CHAP & 182 $(2.25\%)$ &174 (2.14\%)\\
09/F2 &PROC & 7341 $(61.78\%)$ &JRNL & 3677 $(30.95\%)$ &CHAP & 368 $(3.10\%)$ &PAT & 284 $(2.39\%)$ &212 (1.78\%)\\
09/G1 &PROC & 5778 $(57.62\%)$ &JRNL & 3256 $(32.47\%)$ &CHAP & 579 $(5.77\%)$ &MONO & 120 $(1.20\%)$ &294 (2.94\%)\\
09/G2 &JRNL & 6124 $(52.37\%)$ &PROC & 3394 $(29.03\%)$ &ABSP & 723 $(6.18\%)$ &CHAP & 613 $(5.24\%)$ &839 (7.18\%)\\
10/A1 &CHAP & 9208 $(31.89\%)$ &JRNL & 7511 $(26.01\%)$ &PROC & 5722 $(19.81\%)$ &CAT & 1414 $(4.90\%)$ &5023 (17.39\%)\\
10/B1 &CHAP & 5711 $(33.88\%)$ &CAT & 3368 $(19.98\%)$ &JRNL & 2617 $(15.52\%)$ &DICT & 1313 $(7.79\%)$ &3850 (22.83\%)\\
10/C1 &CHAP & 3895 $(30.24\%)$ &JRNL & 3382 $(26.26\%)$ &PROC & 1043 $(8.10\%)$ &CUR & 935 $(7.26\%)$ &3624 (28.14\%)\\
10/D1 &JRNL & 757 $(28.74\%)$ &CHAP & 687 $(26.08\%)$ &PROC & 365 $(13.86\%)$ &REVJ & 310 $(11.77\%)$ &515 (19.55\%)\\
10/D2 &JRNL & 2976 $(34.75\%)$ &CHAP & 1585 $(18.51\%)$ &REVJ & 1203 $(14.05\%)$ &DICT & 730 $(8.53\%)$ &2069 (24.16\%)\\
10/D3 &JRNL & 1283 $(33.52\%)$ &CHAP & 816 $(21.32\%)$ &REVJ & 747 $(19.52\%)$ &MONO & 239 $(6.25\%)$ &742 (19.39\%)\\
10/D4 &JRNL & 2657 $(36.27\%)$ &CHAP & 1736 $(23.70\%)$ &REVJ & 870 $(11.88\%)$ &PROC & 540 $(7.37\%)$ &1523 (20.78\%)\\
10/E1 &JRNL & 1550 $(27.39\%)$ &CHAP & 1254 $(22.16\%)$ &REVJ & 750 $(13.26\%)$ &DICT & 428 $(7.56\%)$ &1676 (29.63\%)\\
10/F1 &JRNL & 2830 $(27.17\%)$ &CHAP & 2642 $(25.37\%)$ &REVJ & 1250 $(12.00\%)$ &PROC & 877 $(8.42\%)$ &2816 (27.04\%)\\
10/F2 &JRNL & 2502 $(30.41\%)$ &CHAP & 1908 $(23.19\%)$ &REVJ & 854 $(10.38\%)$ &MONO & 689 $(8.37\%)$ &2274 (27.65\%)\\
10/F3 &JRNL & 2344 $(27.26\%)$ &CHAP & 2124 $(24.70\%)$ &DICT & 862 $(10.02\%)$ &REVJ & 829 $(9.64\%)$ &2440 (28.38\%)\\
10/G1 &CHAP & 2166 $(30.23\%)$ &JRNL & 1957 $(27.31\%)$ &PROC & 819 $(11.43\%)$ &MONO & 520 $(7.26\%)$ &1704 (23.77\%)\\
10/H1 &CHAP & 1436 $(23.96\%)$ &JRNL & 1434 $(23.92\%)$ &REVJ & 781 $(13.03\%)$ &PROC & 600 $(10.01\%)$ &1743 (29.08\%)\\
10/I1 &JRNL & 1129 $(24.37\%)$ &CHAP & 1128 $(24.35\%)$ &PROC & 574 $(12.39\%)$ &REVJ & 392 $(8.46\%)$ &1409 (30.43\%)\\
10/L1 &CHAP & 3065 $(30.17\%)$ &JRNL & 3014 $(29.67\%)$ &REVJ & 757 $(7.45\%)$ &MONO & 701 $(6.90\%)$ &2622 (25.81\%)\\
10/M1 &CHAP & 1213 $(30.94\%)$ &JRNL & 847 $(21.61\%)$ &REVJ & 383 $(9.77\%)$ &PROC & 300 $(7.65\%)$ &1177 (30.03\%)\\
10/M2 &CHAP & 633 $(29.28\%)$ &JRNL & 526 $(24.33\%)$ &PROC & 234 $(10.82\%)$ &REVJ & 233 $(10.78\%)$ &536 (24.79\%)\\
10/N1 &JRNL & 1742 $(27.19\%)$ &CHAP & 1548 $(24.16\%)$ &REVJ & 625 $(9.76\%)$ &PROC & 605 $(9.44\%)$ &1886 (29.45\%)\\
10/N3 &CHAP & 1200 $(24.46\%)$ &JRNL & 1047 $(21.34\%)$ &DICT & 490 $(9.99\%)$ &REVJ & 475 $(9.68\%)$ &1694 (34.53\%)\\
11/A2 &CHAP & 1800 $(30.29\%)$ &JRNL & 1553 $(26.14\%)$ &DICT & 591 $(9.95\%)$ &REVJ & 531 $(8.94\%)$ &1467 (24.68\%)\\
11/A5 &CHAP & 1264 $(35.43\%)$ &JRNL & 988 $(27.69\%)$ &MONO & 353 $(9.89\%)$ &CUR & 286 $(8.02\%)$ &677 (18.97\%)\\
11/B1 &CHAP & 4145 $(35.32\%)$ &JRNL & 2822 $(24.05\%)$ &PROC & 1558 $(13.28\%)$ &REVJ & 724 $(6.17\%)$ &2487 (21.18\%)\\
11/C1 &JRNL & 2145 $(30.62\%)$ &CHAP & 1596 $(22.78\%)$ &REVJ & 635 $(9.06\%)$ &MONO & 624 $(8.91\%)$ &2005 (28.63\%)\\
11/C3 &CHAP & 1324 $(25.72\%)$ &JRNL & 1318 $(25.60\%)$ &REVJ & 603 $(11.71\%)$ &MONO & 442 $(8.59\%)$ &1461 (28.38\%)\\
11/C5 &JRNL & 2882 $(24.92\%)$ &CHAP & 2687 $(23.23\%)$ &REVJ & 1420 $(12.28\%)$ &DICT & 1338 $(11.57\%)$ &3238 (28.00\%)\\
11/D1 &CHAP & 992 $(32.64\%)$ &JRNL & 907 $(29.85\%)$ &MONO & 311 $(10.23\%)$ &REVJ & 177 $(5.82\%)$ &652 (21.46\%)\\
11/D2 &CHAP & 1109 $(34.91\%)$ &JRNL & 994 $(31.29\%)$ &MONO & 337 $(10.61\%)$ &PROC & 258 $(8.12\%)$ &479 (15.07\%)\\
11/E1 &JRNL & 8406 $(62.19\%)$ &PROC & 1859 $(13.75\%)$ &CHAP & 1216 $(9.00\%)$ &ABSP & 824 $(6.10\%)$ &1211 (8.96\%)\\
11/E2 &JRNL & 1704 $(44.71\%)$ &CHAP & 777 $(20.39\%)$ &PROC & 729 $(19.13\%)$ &ABSP & 162 $(4.25\%)$ &439 (11.52\%)\\
11/E3 &JRNL & 2035 $(50.27\%)$ &CHAP & 960 $(23.72\%)$ &PROC & 493 $(12.18\%)$ &ABSP & 262 $(6.47\%)$ &298 (7.36\%)\\
11/E4 &JRNL & 2473 $(50.88\%)$ &CHAP & 944 $(19.42\%)$ &PROC & 539 $(11.09\%)$ &ABSP & 281 $(5.78\%)$ &623 (12.83\%)\\
12/A1 &CHAP & 1787 $(38.18\%)$ &JRNL & 1494 $(31.92\%)$ &VERD & 641 $(13.69\%)$ &MONO & 344 $(7.35\%)$ &415 (8.86\%)\\
12/B2 &JRNL & 952 $(45.40\%)$ &CHAP & 679 $(32.38\%)$ &VERD & 237 $(11.30\%)$ &MONO & 75 $(3.58\%)$ &154 (7.34\%)\\
12/C1 &JRNL & 1945 $(39.10\%)$ &CHAP & 1788 $(35.95\%)$ &MONO & 284 $(5.71\%)$ &VERD & 280 $(5.63\%)$ &677 (13.61\%)\\
12/C2 &JRNL & 432 $(42.11\%)$ &CHAP & 191 $(18.62\%)$ &MONO & 93 $(9.06\%)$ &PROC & 92 $(8.97\%)$ &218 (21.24\%)\\
12/D1 &JRNL & 1855 $(42.38\%)$ &CHAP & 1398 $(31.94\%)$ &VERD & 429 $(9.80\%)$ &MONO & 229 $(5.23\%)$ &466 (10.65\%)\\
12/D2 &JRNL & 725 $(47.54\%)$ &CHAP & 356 $(23.34\%)$ &VERD & 317 $(20.79\%)$ &MONO & 80 $(5.25\%)$ &47 (3.08\%)\\
12/E1 &JRNL & 1229 $(43.23\%)$ &CHAP & 870 $(30.60\%)$ &MONO & 163 $(5.73\%)$ &VERD & 109 $(3.83\%)$ &472 (16.61\%)\\
12/E2 &JRNL & 1697 $(38.71\%)$ &CHAP & 1390 $(31.71\%)$ &MONO & 250 $(5.70\%)$ &VERD & 215 $(4.90\%)$ &832 (18.98\%)\\
12/E3 &CHAP & 1044 $(38.42\%)$ &JRNL & 969 $(35.66\%)$ &VERD & 310 $(11.41\%)$ &MONO & 166 $(6.11\%)$ &228 (8.40\%)\\
12/F1 &JRNL & 412 $(39.85\%)$ &CHAP & 257 $(24.85\%)$ &VERD & 195 $(18.86\%)$ &MONO & 62 $(6.00\%)$ &108 (10.44\%)\\
12/G1 &CHAP & 748 $(40.06\%)$ &JRNL & 567 $(30.37\%)$ &VERD & 195 $(10.44\%)$ &MONO & 128 $(6.86\%)$ &229 (12.27\%)\\
12/G2 &CHAP & 1080 $(40.00\%)$ &JRNL & 959 $(35.52\%)$ &VERD & 238 $(8.81\%)$ &MONO & 149 $(5.52\%)$ &274 (10.15\%)\\
12/H1 &JRNL & 365 $(35.10\%)$ &CHAP & 272 $(26.15\%)$ &MONO & 128 $(12.31\%)$ &PROC & 71 $(6.83\%)$ &204 (19.61\%)\\
12/H2 &CHAP & 334 $(27.72\%)$ &JRNL & 259 $(21.49\%)$ &REVJ & 146 $(12.12\%)$ &PROC & 137 $(11.37\%)$ &329 (27.30\%)\\
12/H3 &JRNL & 1196 $(37.32\%)$ &CHAP & 793 $(24.74\%)$ &MONO & 266 $(8.30\%)$ &REVJ & 212 $(6.61\%)$ &738 (23.03\%)\\
13/A1 &JRNL & 4600 $(64.85\%)$ &CHAP & 1260 $(17.76\%)$ &OP & 637 $(8.98\%)$ &MONO & 185 $(2.61\%)$ &411 (5.80\%)\\
13/A2 &JRNL & 7127 $(54.52\%)$ &CHAP & 3066 $(23.45\%)$ &OP & 1242 $(9.50\%)$ &PROC & 586 $(4.48\%)$ &1052 (8.05\%)\\
13/A3 &JRNL & 1642 $(58.94\%)$ &CHAP & 574 $(20.60\%)$ &OP & 243 $(8.72\%)$ &PROC & 114 $(4.09\%)$ &213 (7.65\%)\\
13/A4 &JRNL & 3119 $(47.39\%)$ &CHAP & 1688 $(25.65\%)$ &PROC & 607 $(9.22\%)$ &OP & 591 $(8.98\%)$ &576 (8.76\%)\\
13/A5 &JRNL & 759 $(69.57\%)$ &CHAP & 161 $(14.76\%)$ &PROC & 69 $(6.32\%)$ &OP & 69 $(6.32\%)$ &33 (3.03\%)\\
13/B1 &JRNL & 2668 $(35.50\%)$ &CHAP & 2526 $(33.61\%)$ &PROC & 1015 $(13.50\%)$ &MONO & 725 $(9.65\%)$ &582 (7.74\%)\\
13/B2 &JRNL & 2149 $(31.27\%)$ &CHAP & 2061 $(29.99\%)$ &PROC & 1615 $(23.50\%)$ &MONO & 437 $(6.36\%)$ &610 (8.88\%)\\
13/B3 &CHAP & 963 $(34.83\%)$ &PROC & 738 $(26.69\%)$ &JRNL & 708 $(25.61\%)$ &MONO & 148 $(5.35\%)$ &208 (7.52\%)\\
13/B4 &JRNL & 1853 $(40.10\%)$ &CHAP & 1515 $(32.79\%)$ &PROC & 418 $(9.05\%)$ &OP & 332 $(7.18\%)$ &503 (10.88\%)\\
13/B5 &PROC & 971 $(35.43\%)$ &JRNL & 915 $(33.38\%)$ &CHAP & 414 $(15.10\%)$ &ABSP & 193 $(7.04\%)$ &248 (9.05\%)\\
13/C1 &CHAP & 1958 $(35.85\%)$ &JRNL & 1742 $(31.90\%)$ &MONO & 472 $(8.64\%)$ &REVJ & 344 $(6.30\%)$ &945 (17.31\%)\\
13/D1 &JRNL & 2346 $(47.15\%)$ &PROC & 1365 $(27.43\%)$ &CHAP & 616 $(12.38\%)$ &OP & 244 $(4.90\%)$ &405 (8.14\%)\\
13/D2 &JRNL & 1058 $(46.40\%)$ &CHAP & 518 $(22.72\%)$ &PROC & 407 $(17.85\%)$ &OP & 162 $(7.11\%)$ &135 (5.92\%)\\
13/D3 &CHAP & 917 $(34.63\%)$ &JRNL & 824 $(31.12\%)$ &PROC & 412 $(15.56\%)$ &OP & 152 $(5.74\%)$ &343 (12.95\%)\\
13/D4 &JRNL & 1705 $(61.80\%)$ &CHAP & 330 $(11.96\%)$ &PROC & 314 $(11.38\%)$ &ABSP & 159 $(5.76\%)$ &251 (9.10\%)\\
14/A1 &JRNL & 1808 $(34.04\%)$ &CHAP & 1332 $(25.08\%)$ &MONO & 459 $(8.64\%)$ &REVJ & 434 $(8.17\%)$ &1279 (24.07\%)\\
14/A2 &JRNL & 991 $(40.32\%)$ &CHAP & 790 $(32.14\%)$ &MONO & 193 $(7.85\%)$ &REVJ & 161 $(6.55\%)$ &323 (13.14\%)\\
14/B1 &JRNL & 1210 $(27.60\%)$ &CHAP & 1135 $(25.89\%)$ &REVJ & 562 $(12.82\%)$ &MONO & 348 $(7.94\%)$ &1129 (25.75\%)\\
14/B2 &JRNL & 1452 $(30.68\%)$ &CHAP & 1365 $(28.84\%)$ &REVJ & 544 $(11.49\%)$ &MONO & 423 $(8.94\%)$ &949 (20.05\%)\\
14/C2 &CHAP & 1029 $(39.81\%)$ &JRNL & 774 $(29.94\%)$ &MONO & 284 $(10.99\%)$ &CUR & 199 $(7.70\%)$ &299 (11.56\%)\\
14/D1 &CHAP & 1753 $(39.56\%)$ &JRNL & 1388 $(31.32\%)$ &MONO & 385 $(8.69\%)$ &CUR & 257 $(5.80\%)$ &648 (14.63\%)\\\bottomrule
\end{longtable}
\end{scriptsize}

\section{Levenshtein distance}\label{sec:levenshtein}

The Levenshtein distance between two strings (sequences of characters)
is the number of edit operations that are required to transform one
string into the other. The following single-character edit operations
are permitted: (\emph{i})~deletion of a character;
(\emph{ii})~insertion of a character; (\emph{iii})~replacement of a
character with a different one.

Let $S[1..n]$ and $T[1..m]$ be two strings of length $n := |S|$ and $m
:= |T|$, respectively. The Levenshtein distance $L(S, T)$ of $S$ and
$T$ is the value of the auxiliary function $L_{S, T}(n, m)$, where
$L_{S, T}(i, j)$ is defined for all $0 \leq i \leq n$, $0 \leq j \leq
m$ as follows:
\begin{equation}
  L_{S, T}(i, j) := \begin{cases}
    \max\{i, j\} & \mbox{if $i=0$ or $j=0$} \\
    \min\{ L_{S, T}(i-1, j) + 1, L_{S, T}(i, j-1) + 1, L_{S, T}(i-1, j-1) + 1_{S[i] \neq T[j]}\} & \mbox{otherwise}
  \end{cases}\label{eq:lavenshtein}
\end{equation}

\noindent where $1_\mathcal{P}$ if the indicator function, whose value
is $1$ if the predicate $\mathcal{P}$ is true, $0$ otherwise.  $L_{S,
  T}(i, j)$ is the minimum number of edit operations needed to
transform the prefix $S[1..i]$ of $S$ into the prefix $T[1..j]$ of
$T$.  If one of the prefixes is empty ($i=0$ or $j=0$), then the
distance is simply the length of the nonempty prefix.  If both
prefixes are nonempty, $S[1..i]$ can be transformed into $T[1..j]$ by
either:
\begin{enumerate}
\item deleting the character $S[i]$ and transforming $S[1..i-1]$
  into $T[1..j]$; this requires $L_{S, T}(i-1, j) + 1$ edit
  operations;
\item deleting $T[j]$ and transforming $S[1..i]$ into
  $T[1..j-1]$; this requires $L_{S, T}(i, j-1)+1$ edit
  operations;
\item replacing $S[i]$ with $T[j]$ (if they are different,
  otherwise do nothing) and transforming $S[1..i-1]$ into
  $T[1..j-1]$; this require $L_{S, T}(i-1, j-1) + 1_{S[i] =
    T[j]}$ edit operations.
\end{enumerate}

The value $L_{S, T}(n, m)$ can be computed in time $O(nm)$ by
tabulating all values $L_{S, T}(i, j)$ starting from $L_{S, T}(0,
0)$. The Levenshtein distance is zero if and only if $S$ and $T$ are
equal; the maximum value is $\max\{|S|, |T|\}$ when $S$ and $T$
contain distinct sets of characters (e.g., $S=\mbox{``abcdef''}$,
$T=\mbox{``ghijklmnopqrst''}$). The \emph{normalized} Levenshtein
distance $L_n(S, T)$ is defined as:
\begin{equation}
L_n(S, T) := \frac{L_{S,T}(|S|, |T|)}{\max\{|S|,|T|\}}\label{eq:norm-distance}
\end{equation}
\noindent and assumes values in the range $[0, 1]$. 

By definition, a small difference between short documents results in a
larger normalized distance than the same difference between long
documents. Formally, given two pairs of documents $S, T$ and $S', T'$
where $|S| < |S'|,\ |T| < |T'|$ and such that $L_{S, T}(|S|, |T|) =
L_{S', T'}(|S'|, |T'|)$, then according to
Equation~\eqref{eq:norm-distance} we have $L_N(S, T) > L_N(S',
T')$. In short, the same (absolute) difference matters more for short
documents than for long ones.

It is important to remark that the normalized Levenshtein distance
among real-world documents is usually much lower than $1.0$. For
example, the normalized distance between a portion of the United
States Declaration of Independence and a portion of equal length from
the \emph{Divine Comedy} by Italian poet Dante Alighieri is less than
0.8; the distance between two random character sequences is about 0.9.

\end{document}